   \documentclass[12pt,onecolumn,twoside,draft]{IEEEtran} 

\usepackage{multirow}
\usepackage{amsfonts}
\usepackage{epsfig}
\usepackage{amsmath}
\usepackage{amssymb}
\usepackage[nolist]{acronym}
\usepackage[english]{babel}
\usepackage{cite}
\usepackage{color}
\usepackage{stfloats}
\usepackage{algorithm}
\usepackage{algorithmic}
\usepackage{subfigure}

\usepackage{psfrag}
\usepackage{auto-pst-pdf}

\newcommand{\Tdts}{ T_{\text{TS},l} }
\newcommand{\TA}{ T_{\text{A},l} }

\newcommand{\mat}{\pmb}
\newcommand{\TD}{T_{\text{D}}}

\begin{document}

\begin{acronym}
\acro{QoS}{quality-of-service } \acro{MSE}{mean square error }
\acro{MDC}{Multi-description codes} \acro{MRC}{multi-resolution
codes} \acro{RV}{random variable} \acro{GOP}{group of picture}
\acro{RCPC}{rate-compatible punctured convolutional}
\acro{SLV}[PLV]{packet loss visibility} \acro{OFDM}{orthogonal
frequency division multiplexing} \acro{CRC}{cyclic redundancy check}
\acro{EEP}{equal error protection} \acro{UEP}{unequal error
protection} \acro{FEC}{forward error correction} \acro{RB}{resource
block} \acro{CSI}{channel state information} \acro{SNR}
{signal-to-noise ratio} \acro{MCEC}{Motion-Compensated Error
Concealment} \acro{FR}{Full-reference} \acro{VQM}{Video Quality
Metric} \acro{PSAM}{pilot symbol assisted modulation}
\acro{NC}{network coding} \acro{JSCC}{joint source channel coding}
\acro{AWGN}{additive white gaussian noise} \acro{UCSD}{University of
California at San Diego} \acro{UGC}{user-generated content}
\acro{PDA}{Personal Digital Assistant} \acro{TDMA}{Time Division
Multiple Access} \acro{QeA}{question-and-answer} \acro{RD}{rate
distortion} \acro{QoE} {quality of experience} \acro{NC} {network
coding} \acro{MDP}{Markov decision process} \acro{MOS}{mean opinion
score} \acro{PSNR}{peak signal to noise ratio} \acro{RNC}{random
network coding} \acro{RS}{Reed-Solomon} \acro{MD}{multiple
description} \acro{SW}{Slepian-Wolf} \acro{WZ}{Wyner-Ziv}
\acro{SI}{side information} \acro{DSC}{distributed source coding}
\acro{SLEP}{Systematic Lossy Error Protection} \acro{PLR}{packet
loss rate} \acro{PRISM}{Power-efficient, Robust, high  compression,
Syndrome-based Multimedia} \acro{HARQ}{hybrid automatic repeat
request} \acro{MAC}{multiple access control}
\acro{MIMO}{multiple-input multiple-output} \acro{FEP}{forward error
protection} \acro{DP}{dynamic programming} \acro{MV}{multiview}
\acro{RADIO}{RAte DIstortion Optimization} \acro{WMSN}{wireless
multimedia sensor network} \acro{AP}{access point} \acro{DU}{data
unit}\acro{RA}{rate allocation}
\acro{FDMA}{frequency-division multiple access} 
\acro{ISA}{iterative sensitivity adjustment}
\acro{DVC} {distributed video coding}
\acro{DIBR} {depth-image based rendering }
\acro{DAG}{directed acyclic graph}
\end{acronym}

\title{Optimized    Packet Scheduling in Multiview Video Navigation Systems}

\author{ Laura  Toni,~\IEEEmembership{Member,~IEEE},  Thomas Maugey,~\IEEEmembership{Member,~IEEE},   and Pascal Frossard,~\IEEEmembership{Senior Member,~IEEE} \thanks{L. Toni and P. Frossard are with  Ecole Polytechnique F\'ed\'erale de Lausanne (EPFL), Signal Processing Laboratory - LTS4, CH-1015 Lausanne, Switzerland. Email: \{laura.toni, pascal.frossard\}@epfl.ch.    
T. Maugey  is with National Institute for Research in Computer Science and Control, INRIA/IRISA, 35042 Rennes, France. Email: thomas.maugey@inria.fr. }
\thanks{This work was partially funded by the Swiss National Science Foundation (SNSF) under the CHIST- ERA project CONCERT (A Context-Adaptive Content Ecosystem Under Uncertainty), project nr. FNS 20CH21 151569 and  by the Hasler Foundation under the project NORIA (Novel image representation for future interactive multiview systems).}
}
\maketitle
\thispagestyle{empty}
\begin{abstract}
In multiview video systems, multiple cameras generally acquire the same scene from different perspectives, such that users have the possibility to select their preferred viewpoint. This results in large amounts of highly redundant data, which needs to be properly handled during encoding and transmission over resource-constrained channels. In this work, we study coding and transmission strategies in multicamera systems, where correlated sources send data  through a bottleneck channel to a central server, which eventually transmits views to different interactive users.  We propose a dynamic correlation-aware packet scheduling optimization under delay, bandwidth, and interactivity constraints. The optimization relies both on a novel rate-distortion model, which captures the importance of each view in the 3D scene reconstruction, and on an objective function that optimizes resources based on a client navigation model. The latter takes into account   the   distortion   experienced by interactive clients as well as the distortion variations that might be observed by     clients during multiview navigation.  
We  solve the scheduling problem with a novel trellis-based solution, which permits  to formally decompose the multivariate optimization problem  thereby significantly reducing the computation complexity. Simulation results show the gain of the proposed algorithm compared to baseline scheduling policies.
More in details, we show the gain offered by our dynamic scheduling policy compared to static camera allocation strategies and  to schemes with constant coding strategies. Finally, we show that the best scheduling policy consistently adapts to the most likely user navigation path  and that it minimizes distortion variations that can be very disturbing for users in traditional navigation systems.
\end{abstract}

\section{Introduction}
The bursting diffusion of novel video sharing and streaming applications has recently  opened the   era of user-centric multimedia.  In new multimedia services, users do not passively download media content,  but rather dynamically select the content they are interested in.  Resource allocation strategies  cannot   anymore be  built offline, according to predefined users behaviors.  Hence, effective real-time interactive services can only be devised if adaptivity to channel conditions and users dynamics represents the primary feature  of media delivery strategies. 

In order to accomodate for dynamic   networks, online resource allocation strategies have been proposed for video applications \cite{ChoMia:J06,FuSchaar:J12}.  However,    few works have extended the study to interactive multiview  video streaming applications.  The main challenges with these new applications are the proper handling of  the spatial  correlation that exists  among  different camera views capturing the same 3D scene, and   the uncertainty of users requests since those can freely navigate in the multiview content. These two challenges have not been addressed together, to the best of our knowledge.    Spatial correlation has been taken into account in multisource resource allocation strategies for sensor networks \cite{WanDaiAky:C11} and for more general wireless networks    \cite{Chakareski:J14,Toni:J14}.  
Interactivity of   users is however mostly overlooked in resource allocation solutions in the literature.  In this work, we exactly aim at filling this gap by proposing resource allocation strategies where users' navigation features play a key role in the  optimized scheduling of information from correlated sources.

We consider a live acquisition scenario  in which multiple cameras acquire frames of the same scene but from different perspectives.  Each camera acquires the scene, produces and stores frames in its buffer, possibly in different  independently encoded versions.   We assume that no content information is exchanged among cameras due to the system configuration or resource limitations. 
 The only minimal  information that is known a priori is the position of the cameras, which is possibly updated when cameras change positions in dynamic settings. 
The encoded  frames are then sent from the cameras to a central server through a bottleneck channel
under deadline constraints imposed by the streaming application. A server gathers the camera frames and eventually serves the clients requests. Our objective in such a system is to maximize the temporal quality variations for users navigating in the multiview content.  In particular, when  
the channel constraints do not permit to send all captured views, it becomes important to  optimize  the scheduling policy in such a way that the quality in the reconstruction of the multi-camera data is maximized and both bandwidth and time constraints are met.  

 \begin{figure}[t]
\begin{center}
\includegraphics[width=0.9\linewidth,  draft=false]{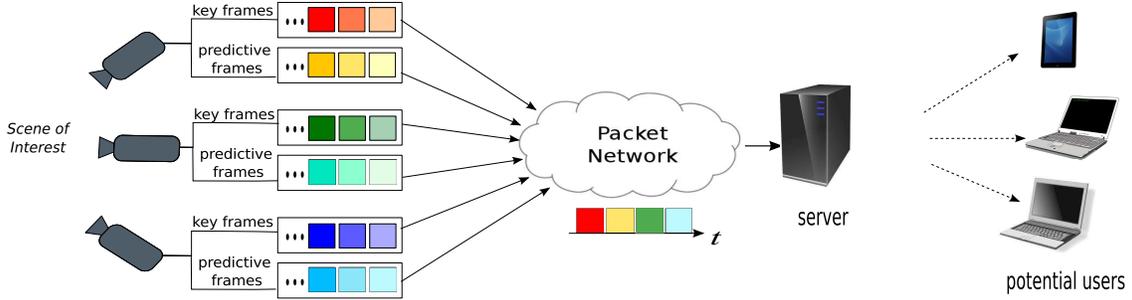}
\caption{Multicamera system with  bottleneck network.  Each camera acquires, encodes, and temporally stores frames of a given view of  the 3D scene. Frames are sent  through a bottleneck channel to a central server, that eventually serves clients' requests.}\label{fig:Scenario_PF}
\end{center}
\end{figure}
%

We propose   a new  \emph{navigation-aware} packet scheduling algorithm for streaming from multiple correlated cameras  in bandwidth-limited networks.  We consider a correlation-based \ac{RD} model that is specific to multi-camera systems and we formulate a packet scheduling optimization problem that minimizes   the distortion of the data available at the server, while also reducing the   distortion variations along most likely navigation paths. We further propose to select the coding structure   dynamically  according to the packet scheduling strategy. In this way,  we  are able to constantly adapt the set of coded packets that are transmitted to   the server to the channel conditions,  to the content information, as well as to the expected users behavior. To solve the resulting multivariate optimization problem,  we propose a \emph{novel solving algorithm} that is able to reduce the computational complexity of the scheduling solution by  decomposition while preserving its optimality.  Simulation results demonstrate that our new dynamic scheduling algorithm outperforms baseline scheduling policies with static  coding strategy and   transmission schemes with limited  adaptivity.     In particular, we show that   information about users' interaction in the problem formulation leads to an improvement in terms of perceived quality with respect to classical scheduling algorithms. Simulation results also outline the limitations of commonly used static encoding strategies that have poor performance   in highly constrained scenarios. Finally, the results show that smooth quality variations  are  experienced over the navigation path with our optimal scheduling strategy, which is not the case of state-of-the-art  scheduling solutions that merely target minimal average distortion.
 

 The remainder of this paper is organized as follows. Related works on multiview video streaming are described in Section \ref{sec:related_works}.  Section \ref{sec:frameworks} describes the multicamera system, together with 
our  new rate-distortion function for the representation of 3D scenes.  The packet scheduling problem is formulated in Section \ref{sec:packet_scheduling} and the trellis-based optimization solution is provided in Section \ref{sec:solving_method}. In Section \ref{sec:results}, we discuss the simulation results, and we conclude in Section \ref{sec:conclusion}.

\section{Related Works}
\label{sec:related_works}
Although resource allocation strategies have been widely investigated in the literature for single view video streaming, there are still many open challenges in multiview video  scenarios. In this section, we describe the works related to multiview scheduling policies and highlight the lack of complete solutions that   take  into account both  source correlation in multiview settings and users interactivity in navigation applications.

Several works have studied the problem of scheduling of correlated video sources \cite{WanDaiAky:C11,  Dieber:J11, YuSharma:J10, Chakareski:J14,Toni:J14}. The work  in \cite{WanDaiAky:C11}   proposes a spatial correlation model for visual information in \acp{WMSN} and introduces an entropy-based analytical framework to evaluate the visual information offered by  multiple cameras. 
The system however only solves a static correlation-based   camera selection  problem, while  we consider a dynamic correlation-based packet scheduling optimization problem in our work.  More dynamic camera scheduling for \acp{WMSN} have been proposed in  \cite{Dieber:J11, YuSharma:J10}, where  optimal resource allocation strategies adapt to the dynamics of the system. The  optimization however mainly addresses surveillance networks or object tracking   scenarios, where the  problem formulation consists in maximizing the coverage of the area monitored by the camera sensors while preserving the life time of the network. In our work, we rather optimize the experienced quality of interactive users and the expected quality variations perceived over likely navigation paths.  Other works \cite{Chakareski:J14,Toni:J14} have studied the problem of source correlation aware  transmission policy optimization for multiview scheduling. However, the interactivity of users has been neglected and only predefined coding strategies have been considered so far.

Some   prior  studies    address  the problem of providing to users interactivity in selecting views, while saving on transmitted bandwidth and view-switching delay~\cite{KurCivTek:J07, CheVelOrt:J11, LiuQinMaZha:J10, CheOrtChe:J11,KueSik:C08,CheZhaSunShi:C09}. The work in~\cite{CheOrtChe:J11} is mainly focused on coding views with a minimum level of redundancy in order to simplify the view switching, and the works in~\cite{LiuQinMaZha:J10,PanIkuBanWat:C11}   optimize  the selection of views to be encoded and transmitted based on the user interest. The authors in~\cite{KueSik:C08,LouGuaHuaJia:J07}   investigate  the transmission of  multiview video coded streams on P2P networks and IP multicast, respectively. These works mainly focus on the coding optimization  proposed as an a priori defined solution  to provide interactive access to the different views \cite{CheOrtChe:C09}. In our work, we rather dynamically optimize the coding modes and the scheduling   of video frames for interactive multiview navigation.  We extend our preliminary work in \cite{Toni:VCIP14} to include users' interactivity in the scheduling optimization.  This allows the system to dynamically adapt the transmission of the coded frames to various system's dynamics and to outperform the above  transmission policies of a priori encoded frames.

Finally, in our scheduling optimization we aim at minimizing the experienced distortion as well as the   temporal variations of the experienced distortion for interactive users. Recently, it has been shown the importance of studying the temporal quality variations in adaptive streaming strategies  \cite{Li:2014,Chen:J14}. These works  target  single view adaptive streaming over HTTP. We follow similar intuitions and extend the   mixed objective function composed on both the perceived distortion and the temporal distortion variation to multiview video navigation applications.

\section{Multiview Acquisition Framework}
\label{sec:frameworks}
In the following, we first  
 present  the multicamera system  considered in our work.  
Then, we describe in details  the adopted  coding scheme   and show that the correlation between the different cameras plays a crucial role in the reconstruction of images in multiview navigation in resource constrained environments.  Finally, we propose a new rate-distortion model for the representation of the 3D scene information.

\subsection{Multi-camera acquisition system}
We consider a system with $M$ cameras that acquire images and depth information of a 3D scene from different viewpoints.  Each frame can be encoded as a key frame (i.e., as intra-coded frame) or   dependent  frames, 
that  are indepdentely coded with \ac{DSC} techniques  using correlated key frames  as \ac{SI}. 
We denote by P the dependent frames that use   key frames correlated in the temporal domain as possible SI, while WZ frames use  neighboring key frames both in the temporal and the spatial domains. 
The encoded versions are stored in each camera buffer for a maximum of $\TD$ time slots, where  $\TD$ represents the frame deadline. The coded frames are then transmitted to a central server, which eventually serves the requests from interactive users.

Each camera acquires temporally consecutive
frames, which are correlated, especially for static or low-motion 3D scenes: this is the \emph{temporal correlation}   in image sequences. Neighboring cameras might also acquire  overlapping portions of the same scene; this   leads to correlated  frames due to  the \emph{spatial correlation} between multiview cameras.  We assume that no information is exchanged between camera  except   minimal information about the camera position. With this position information,  each camera is able to coarsely  estimate  the contribution that it can offer to the reconstruction of neighbor views  \cite{Toni:J14}.  For an image $F$, we denote by $\rho(F| \mathcal{F})$ the level of correlation between $F$ and its neighbors (either in time or space) in a set $\mathcal{F}$. This level $\rho(F| \mathcal{F})$ represents the proportion of the image $F$ that can be estimated from   $\mathcal{F}$. In the case of only one   image $F^{\prime}$  composing the set $\mathcal{F}$, we have $\rho(F| F^{\prime})$ as the correlation level between the two images.

In practice,  network limitations might prevent the transmission of all views to the server, which eventually serves users according to their different requests  while  navigating in the multiview dataset.  At the decoder side, missing images can be  reconstructed from correlated neighboring views if available  \footnote{The decoding process can be physically performed either at the central server or at the clients. Our problem formulation is general enough to consider both cases. }.    Both temporal and spatial correlation  might help in reconstructing missing images, so that   it is important to accurately select the images   to be transmitted and  their encoding mode (i.e., key-frame or  dependent frame), such that the average distortion is minimized and  quality variations along the users' navigation paths are limited. This is precisely the frame scheduling problem considered in this paper. Before formulating the problem more precisely, we provide below details about the coding modes and the rate-distortion model used in our multiview system.

\subsection{Frame Coding Modes}
We now give some details on the encoding and decoding structure that are considered in our interactive multiview system. 
At the  camera side, each frame is independently encoded as a key frame. It is also encoded  as  P or WZ frames, with no a priori information on  the frames that will be available at the decoder side (i.e., with no a priori knowledge of the actual frame scheduling).   
For each image $F$, we define a neighborhood  as the images that can be used as side information for decoding the P or WZ version of $F$. Ideally, the neighborhood of image $F$ should include all images correlated to $F$. This would increase the chances for $F$ to be decoded if transmitted as  dependent frame.  Similarly to \cite{Cheung_G_2011},  each P or WZ frame is encoded  with respect to the least correlated frame in the neighborhood, that is with a coding rate that guarantees decoding in the worst-case scenario. The
lower is the correlation between $F$ and the least correlated image in the  neighborhood, the less efficient is the coding of $F$ as  dependent frame.  For this reason, the neighborhood is limited  to    any frame that has a level of correlation with  $F$ greater than a predefined threshold value $\beta$.   
   
More in details,   the $m$-th camera   acquires the frame $F_{t,m}$ at time  $t$.  For any acquired  frame $F_{t,m}$, we define the set of possible \ac{SI} frames in spatial and temporal domain respectively as
\begin{align}\label{eq:neighborhood}
\mathcal{N}_S(F_{t,m}) &=\left\{ F_{t,l} \text{ s.t. }  \rho(F_{t,m}|F_{t,l})  > \beta_S,   \text{ with }  \ l\in [1,M] \right\} 
\\  \nonumber
\mathcal{N}_T(F_{t,m})&=\left\{ F_{t^{\prime},m} \text{ s.t. }  \rho(F_{t,m}|F_{t^{\prime},m})  > \beta_T,   \text{ with }  t^{\prime}\leq t  \right\} \,.
\end{align}
The P version of $F_{t,m}$ is  encoded  considering as SI only neighbor frames   in the temporal domain, i.e., $\mathcal{N}_T(F_{t,m})$. 
 Analogously, we assume that the WZ version of $F_{t,m}$ is  encoded assuming a SI region, which extends in both time and space, and it defined as  
$\mathcal{N}(F_{t,m}) =  \{\mathcal{N}_S(F_{t,m})  \cup  \mathcal{N}_T(F_{t,m}) \}$.  
Note that only key frames within the defined neighborhoods can be used at the decoder as SI.

We assume that  the  WZ version of the frame $F_{t,m}$, which  has been encoded by considering  $F_{t,l}$ as side information,   has an encoding rate  of $\mathcal{R}(F_{t,m}|F_{t,l})=\left[ 1-  \rho(F_{t,m}|F_{t,l}) \right] R_{t,l}^K $, where $R_{t,l}^K$ is the encoding rate of the key version of $F_{t,l}$. Thus, since encoding is based on worst case SI frames,  each WZ and P frame is encoded at  a rate of 
\begin{align}\label{eq:WZrate}
 R_{t,m}^{WZ} & = \max_{F_{t^{\prime},l}Ê\in \mathcal{N}(F_{t,m}) } \left\{ \left[ 1-  \rho(F_{t,m}|F_{t^{\prime},l}) \right] R_{t^{\prime},l}^K \right\}  
   \\  \nonumber
 R_{t,m}^{P} & = \max_{F_{t^{\prime},m}Ê\in \mathcal{N}_T(F_{t,m})} \left\{ \left[ 1-  \rho(F_{t,m}|F_{t^{\prime},m}) \right] R_{t^{\prime},m}^K \right\} \,.
\end{align}
Note that a more scalable scheme can be considered by assuming different WZ or P versions, each one with different  $\beta$ thresholds and thus different encoding rates. This would refine the optimal scheduling solution, but it would not change our  problem formulation. For the sake of simplicity,  we consider one WZ and P version per frame in the following.

At the  receiver side, each received  key  frame is decoded independently. We denote by $\mat{\chi}$ the key frames   available at the decoder. Key frames in  $\mat{\chi}$ and in the  neighborhoods $\mathcal{N}(F_{t,m})$ or $\mathcal{N}_T(F_{t,m})$ are used to decode WZ or P frames, respectively. 
%
The missing images that have not been transmitted at the server are  estimated, at the decoder, with   view interpolation algorithms using information from neighbor  key frames.  The neighborhood of a missing image $F_{t,m}$ is given by the set of key frames with a non null correlation  with $F_{t,m}$. 
More precisely, a missing  view is recovered  from  the neighbor key frames available at the receiver   by  \ac{DIBR} techniques~\cite{Fehn_C_2004}. Typically,  \ac{DIBR} algorithms use  depth  information in order to  estimate by projection the position of pixels from view $k$ in  the missing view $n$. The projected pixels are generally of  good precision (depending on the accuracy of the depth map
\cite{Muller_K_2011_pieee_tdv_rudm}) but they do not cover the whole
estimated image, due to visual occlusions. 
 The portion of the image $F_{t,m}$ that can be recovered  (i.e., not occluded)  by the neighbor frames   is $\rho(F_{t,m}|\chi)$. The remaining occluded pixels  covers a portion  $1-\rho(F_{t,m}|\chi)$ of the image $F_{t,m}$ and are recovered by  inpainting techniques~\cite{Daribo_I_2010}.

\subsection{Navigation-Aware Rate-Distortion Model}
\label{sec:RD}


We now   propose a novel rate-distortion model   for our multiview video navigation   framework.   
Recall that   only a  subset of the compressed images captured by all cameras is     transmitted to the server, which should be able to serve   any client requests. This is equivalent to offer to the client the possibility to efficiently reconstruct any camera view at any time instant.   If  the frame $F_{t,m}^K$ (i.e., the key-frame) is  available at the decoder, the   distortion is directly dependent on  the source rate $R^{{K}}_{t,m}$. The   distortion function is evaluated from the general  expression of the \ac{RD}
function of an intra-coded frame  with   high-rate
assumption~\cite{CovTho:B91}:
\begin{equation}\label{eq:RD_3}
d(R^{{K}}_{t,m})= \mu_I \sigma_I^2 \, 2^{-2   R^{{K}}_{t,m} }
\end{equation}
where   $\sigma^2_I$ is the spatial variance of the frame   and $\mu_I$ is a constant
depending on the source distribution.  This model has been chosen because   it is quite simple and yet accurate. However,  our packet scheduling framework is general and other source rate-distortion functions could be used.
 We assume that all key frames are encoded at the same rate, i.e., $R_{t,m}^{{K}}=R^K, \forall \{t,m\}$, to target    an almost constant quality of the scene across space and time, in such a way that a smooth interactive system can be offered to the user in ideal conditions. This translates to having the  encoding rate for all key views when image content is similar in different views.  The model presented in the following can however be easily extended to a multi-rate encoding system.
 
If the key version of $F_{t,m}$ is missing at decoder but a WZ or P versions are available,  the frame $F_{t,m}$ is   reconstructed  also at the distortion  $d(R^K)$ as long as their rate has been chosen accordingly to Eq. \eqref{eq:WZrate} and side information is available.  In the remaining case in which neither the key nor the WZ or P versions of $F_{t,m}$ is received, this frame is reconstructed through \ac{DIBR} using   the key   frames available at the decoder, as explained above. The part of the image that can be reconstructed from neighbor frames, i.e., $\rho(F_{t,m}| \mat{\chi})$, has a distortion equal to the distortion of key frames, namely,  $d\left(R^{K} \right)$. The remaining part corresponding to occlusions    is recovered with inpainting techniques at a distortion $d_{\text{max}}$. This results in an overall distortion of the reconstructed image given of $\rho(F_{t,m}| \mat{\chi} )  \cdot d\left({R}^{K}\right) + (1-\rho(F_{t,m}| \mat{\chi} ) )d_{\text{max}}$. 

 We denote by the operator $\mathcal{I}(F)=1$ the availability of frame $F$ at the decoder, and by  $\mathcal{I}(F)=0$ its absence. The frame $F$ is either the key version $F_{t,m}^K$ of $F_{t,m}$, its WZ version $F_{t,m}^{WZ}$ or its P version $F_{t,m}^P$. 
Finally, we can write the distortion of frame $F_{t,m}$   at decoder   as
\begin{align}\label{eq:RD_final_2}
& D_{t,m} (R^K|  \mat{\chi}) = \\ \nonumber
& \left\{  
\begin{array}{ll}
\mu_I \sigma_I^2 \, 2^{-2   R^K   }   &\hspace{-0.5cm} \hspace{.3cm} \text{ if }  \mathcal{I}(F_{t,m}^K) = 1 \\
 &\hspace{-0.5cm} \text{or if }  \mathcal{I}(F_{t,m}^{WZ})=1  \text{  and }   \sum_{F_{t^{\prime},l}\in  \mathcal{N}(F_{t,m})}   \mathcal{I}(F_{t^{\prime},l}^K) \geq 1\\
 &\hspace{-0.5cm} \text{or if }   \mathcal{I}(F_{t,m}^P) =1  \text{  and }   \sum_{F_{t^{\prime},m}\in  \mathcal{N}_T(F_{t,m})}   \mathcal{I}(F_{t^{\prime},m}^K)    \geq 1\\
\rho(F_{t,m}| \mat{\chi} )  \cdot \mu_I \sigma_I^2 \, 2^{-2   R^K   } + (1-\rho(F_{t,m}| \mat{\chi} ) ) \cdot d_{\text{max}}&  \text{otherwise.}  
\end{array}
\right.   
\end{align}
Note that the overall distortion does not depend  on the   coding rates of the P and WZ frames, as those are set in a conservative way according to Eq. \eqref{eq:WZrate}.

 The interactivity offered to clients is captured by  the camera popularity $P_l$,   the portion of   clients that can request the view $F_l$. 
Each encoded frame $F_{t,m}$ has a  popularity   $P_{t,m}$, with  $\sum_m P_{t,m}=1$, which it is defined as the probability that an interactive user requests  frame $F_{t,m}$.    Furthermore,  the probability for a user to navigate from frame $F_{t,m}$ to frame $F_{t+1,l}$ is denoted by $w_{m,l}^t$, with $\sum_l w_{m,l}^t=1$. The expected distortion experienced by interactive users navigating in the 3D scene acquired at time $t$ is given by  
\begin{align}\label{eq:exp_distortion} 
\sum_{m=1}^M P_{t,m} D_{t,m} (R^K|  \mat{\chi}) \,.
\end{align}
Beyond the popularity-weighted distortion, another important metric in 3D interactive services is  the smoothness of the navigation, i.e., the quality variation experienced during the navigation. Varying quality while changing  view can result in an annoying degradation in quality of experience. The smoothness of the navigation is given by  
\begin{align}\label{eq:smoothness}
 \sum_{m=1}^M   \sum_{l=1}^M   w_{m,l}^t P_{t-1,l} \left|  D_{t-1,l} (R^K|  \mat{\chi})  -  D_{t,m} (R^K|  \mat{\chi})   \right|\,.  
\end{align}
It is worth noting that,   we consider a novel rate-distortion model for interactive navigation, which is able to combine the overall distortion from Eq. \eqref{eq:exp_distortion} and the smoothness experienced by users while navigating, given by Eq.  \eqref{eq:smoothness}.

\section{Packet Scheduling Optimization}
\label{sec:packet_scheduling}
We now describe the problem of   rate-distortion optimal packet scheduling  for multiview camera systems. First, we describe the transmission process considered in our work, then we propose a new problem formulation based on the rate-distortion model described above.

\subsection{Transmission policy}
\label{Transmission policy}

Each image acquired at a given time instant from a particular camera  
is packetized into multiple data units (DUs) (one per encoded
version), and stored in the camera buffer.  The DUs representing the key versions contain
texture and depth information about the 3D scene, while WZ
or P versions   only contain the encoded texture information,
since they are not used to reconstruct missing views. We consider a channel with successive time slots $\tau$, each one of duration $\Delta\tau$ and each one being a transmission opportunity.  At each $\tau$, the scheduler decides the best set of DUs to schedule, that is the set of DUs that will optimize the navigation of the users while satisfying bandwidth constraints.  
 Lossless transmissions are considered, such that scheduled packets
are eventually available at the server.
   Let represent each image $F_{t,m}$ by a generic image $F_l$, where we have dropped the subscript $(t,m)$ in favor of   a   general subscript $l$, for the sake of clarity. The image $F_l$
 is  acquired at time  $\TA$ and  expires at time $\Tdts$.    We then define the set of candidates for being sent at time  $\tau$ as the set of acquired  images that do not expire before the transmission is completed, i.e.,  $\mathcal{L}=\{F_l \text{ s.t. } \TA\leq \tau, \tau + \Delta\tau < \Tdts \}$.Ê 
 The different encoded versions of views in $\mathcal{L}$ are candidate DUs for being scheduled. However, we impose the following scheduling policies: i)  only one DU among WZ, P, and key versions of the same image can be scheduled; ii)   a WZ or P version is scheduled  only if some \ac{SI} image has already been scheduled. Finally, since both the channel conditions and content models may vary over time, leading to different scheduling policies at different transmission opportunities,   the   scheduling policy is  updated periodically  at each new transmission opportunity $\tau$.

\subsection{Problem Formulation} 
The objective is now to select the best transmission policy, in order to minimize the distortion and the distortion variations under channel constraints, content dynamics, and client interactivity behavior. 
We define a scheduling policy at time $\tau$ as   $\mat{\pi} =[ \mat{\pi}_{1} ,  \mat{\pi}_{2}, \ldots,  \mat{\pi}_{|\mathcal{L}|}]^T$ where $\mat{\pi}_{l} = [\pi_{l,1}, \pi_{l,2}, \pi_{l,3}]$, and $\pi_{l,1}, \pi_{l,2}, \pi_{l,3}$ are the scheduling policy of respectively  the key, WZ, and the P DU of $F_l$.   \footnote{ $\mat{\pi}$  depends on the time     $\tau$ at which the policy is optimized but, for sake of clarity,   we omit this dependency in the notation.}
A policy binary $\pi_{l,i}$ defines transmission of the key, WZ, and the P DU of $F_l$, for $i=1,2,$ and $3$, respectively.   In other words,  $\pi_{l,i}=1$ means that the associated DU is sent at the current  transmission opportunity $\tau$.   
We can then express our optimization problem as follows
\begin{subequations}\label{eq:PF}
\begin{align}     
\text{\bf Problem 1:} \nonumber \\
 \min_{\mat{\pi}} & \, \overline{D}_{\mat{\pi}} = \sum_{l:  \TA\leq\tau\leq \Tdts}   P_l D_l(\mat{\pi}| \mat{\chi} )   
 + \lambda \left\{ \sum_{j:  T_{\text{A},j}= \TA -1}   w_{jl} P_j \left| D_j(\mat{\pi}| \mat{\chi} )   -D_l(\mat{\pi}| \mat{\chi} )    \right| \right\}
\label{eq:PF_1}   \\
\text{s.t. } &  \sum_l \pi_{l,1} R^{(K)}_l + \pi_{l,2} R^{(WZ)}_l + \pi_{l,r} R^{(P)}_l \leq C_{\tau}  \label{eq:PF_2}  \\
& \sum_i \pi_{l,i}   \leq 1, \ \ \forall l  \label{eq:PF_1} \\
&  {\pi}_{l,2}^T \leq  \sum_{F_{m} \in  \mathcal{N}(F_{l})}  \pi_{m,1}    \label{eq:PF_3} 
 \\
&  {\pi}_{l,3}^T \leq  \sum_{F_{m} \in  \mathcal{N}_T(F_{l})}  \pi_{m,1}    \label{eq:PF_4} 
\end{align}
\end{subequations}
 where the objective function is composed of the expected distortion, defined in Eq. \eqref{eq:exp_distortion}, and the smoothness of the navigation,  defined in Eq. \eqref{eq:smoothness},  experienced by interactive users while navigating the 3D scene.   If $\pi_{l}=1$ or $F_l\in  \mat{\chi}$, then  $\mathcal{I}(F_l)=1$, where $\mathcal{I}(F_l)$ is the availability  of frame $F_l$ at the decoder.      We have denoted by $\lambda$ the multiplier that allows to assign the appropriate weight to quality variations in the objective metric, as already adopted in similar optimization problems \cite{Chen:J14}.  Eq.  \eqref{eq:PF_2} imposes the bandwidth constraint due to the network conditions at the current transmission opportunity, Eq. \eqref{eq:PF_1} imposes that at most one encoded version of  an image is scheduled, and 
 Eq.  \eqref{eq:PF_3} and  Eq.  \eqref{eq:PF_4} force a  dependent frame to be scheduled if and only if at least one side information key frame is available at the decoder.  Finally, $\mat{\chi} $ is the set of DUs already available at the decoder side and it represents the results of past scheduled decisions. 

\section{Trellis-Based Scheduling Algorithm}
\label{sec:solving_method}
The  above scheduling optimization  problem is  challenging due to     the 
 inter-dependency  and the redundancy that subsist  among candidate DUs. The \emph{coding-dependence}   is imposed by the coding structure and it is such that  a WZ or P  frame can be decoded only if at least one side information key frame can also be decoded. The \emph{reward-dependence} is rather coming from the correlation among neighboring key frames. Since a scheduled key frame can  reconstruct missing frames,  the exact reward of scheduling  a key DU is not known a priori, but it depends on   the scheduling policy of the correlated DUs.

Because of coding- and reward-dependence, the    optimization in Eq.  \eqref{eq:PF}  cannot be solved by  conventional optimization frameworks.   
Solutions proposed  in \cite{ChoMia:J06,FuSchaar:J12} could be adopted  in the case of coding-dependence,  but they do not address  the reward-dependence.  Although  a formal scheduling optimization has been posed for redundant DUs in \cite{Ort:J09}, computational complexity remains an open issue.   A viable solution for  \emph{reward-dependent} DUs is the trellis-based algorithm  proposed in \cite{Toni:J14}, where  branches   in the trellis are pruned to   reduce the complexity. 
However, this pruning applies only among key frames DUs and not among key and  dependent candidate frames that are considered in this work.   Thus, the solving method to optimize the scheduling policy in multiview systems is still a very challenging problem.  

  Here, we propose a trellis-based solution  that allows to   reach    \emph{optimality} while reducing at the same time the computational complexity of a complex full search solution.  The heterogeneity of the DUs  enables us to include our scheduling rules in the  construction of the trellis. These rules  provide    an elegant structure to decouple reward-dependent DUs (key frames) from the reward-independent ones (dependent frames), thereby significantly reducing the computation complexity.

\begin{figure}[t]
\begin{center}
\includegraphics[width=.9\linewidth,   draft=false]{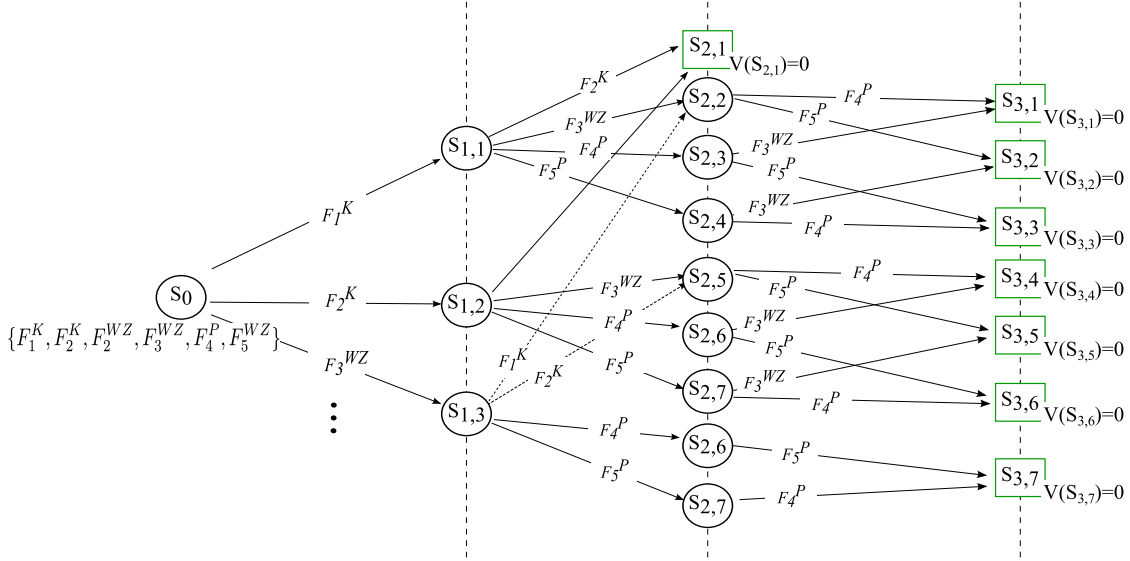}
\caption{  Example of the trellis construction at time $\tau$ in which the initial set of candidates at the initial state  $S_0$   is $\mathcal{A}(S_0)=\{F_1^K, F_2^K, F_2^{WZ}, F_3^{WZ}, F_4^P, F_5^P\}$. The channel bandwidth allows to schedule two key frames or one key frame  and two WZ or P frames.   Black circle nodes denote states with non-null set of candidates or non-zero remaining channel bandwidth, while green square nodes represent final states  from which no further action is  taken.   The frame label provided on each branch going from $S_{i-1,y}$ to $S_{i,x}$ indicates the action taken from $S_{i-1,y}$ that leads to $S_{i,x}$. }\label{fig:Full_trellis}
\end{center}
\end{figure}

\subsection{Trellis Construction}
We start from an initial state $S_0$, characterized by the initial set of candidate DUs. We then construct a trellis,  as depicted in Fig. \ref{fig:Full_trellis}, where each branch is an action (i.e., the scheduling of a DU). Each action $a$ has a cost given by the size of the scheduled DU  and a reward in terms of  distortion gain $\delta(a)$, derived as the difference in the objective  function  $\overline{D}_{\mat{\pi}}$ in Eq. \eqref{eq:PF_1}    with and without the DU corresponding to the scheduling action $a$. 
Each node in the trellis is a state. The state $S_{i,k}$  is the $k$-th node corresponding to the $i$-th DU that has been  scheduled. It  is defined by the set of feasible actions  that can be taken at  node $S_{i,k}$ (i.e., set of possible DUs  to schedule   at $S_{i,k}$) $\mathcal{A}(S_{i,k})$ and by the remaining channel bandwidth $C(S_{i,k})$, evaluated as  the channel bandwidth $C_{\tau}$ minus the sum of the transmission costs corresponding to  the decisions taken along the path  from $S_0$ to $S_{i,k}$.  Note that  $\mathcal{A}(S_{i,k}) = \mathcal{A}^p(S_{i,k}) \cup \mathcal{A}^k(S_{i,k})$, where $\mathcal{A}^p(S_{i,k})$ and $\mathcal{A}^k(S_{i,k})$ are the set of  dependent and key candidate DUs, respectively.   
An action $a\in \mathcal{A}(S_{i,k})$ taken from the state $S_{i,k}$ leads to a successor state $S_{i+1,j}$.  The DU scheduled by $a$ is removed from the set of candidates DUs for the future states.   In the successor states, also DUs corresponding to the same image but at different encoding versions are removed from the set of candidate DUs, to respect the constraint provided in Eq. \eqref{eq:PF_1} of Problem 1. 
 We denote by    $P(S_{i+1,{k^{\prime}}}|S_{i,k}, a)$   the probability of arriving in state $S_{i+1,{k^{\prime}}}$ by taking action $a$ from state $S_{i,k}$. In our case, given the action $a$ the future state is deterministically evaluated by the remaining  candidate DUs and the channel bandwidth.  This means that among all future states, only one will be such that $P\left(S_{i+1,{k^{\prime}}}|S_{i,k}, a\right) =1$ and $0$ for the remaining states.

Each state $S_{i,k}$ is further  characterized by  the value function $V_{\pi}(S_{i,k})$ under a scheduling policy $\pi$, which   represents  the reward when starting from state $S_{i,k}$ and following  the policy $\pi$ thereafter. In our problem,  $\pi$ is the set of actions taken from  $S_{i,k}$ and thereafter. If at state $S_{i,k}$ the remaining channel bandwidth is zero or the set of candidates is null,  $S_{i,k}$ is a final state,  and no further actions can be taken. The value function for a final state  is always null, i.e.,  $V_{\pi}(S_{j,k})=0$~\cite{sutton1998}.     
Finally, the full-path  going from $S_0$ to a final state,   which leads to the maximum total reward, corresponds to  the best set of DUs to be scheduled. From   the Bellman's optimality equations,   the best full-path  can be found by backward induction  from every final state $S_{i,k}$  as follows \cite{cormen2001}
\begin{equation}\label{eq:Bellman}
V_{\pi^{\star}}(S_{i,k}) = \max_{a\in \mathcal{A}(S_{i,k})}\left\{ \delta(a) + \sum_{k^{\prime}} V_{\pi^{\star}}(S_{i+1,{k^{\prime}}}) P\left(S_{i+1,{k^{\prime}}}|S_{i,k}, a\right)  \right\} \,.
\end{equation} 
Such a  problem however  suffers from large computational complexity, namely the trellis construction is exponentially complex.  In order to reduce the complexity, we impose  the following two rules in the trellis construction.
\vspace{2mm}  \newline
 \emph{ {\bf Rule 1:} }  If the action $a$ corresponds to the scheduling of a  dependent frame, then key frames cannot be scheduled in any successor state.
\vspace{2mm}  \newline
The first  rule avoids to construct redundant  paths with the same reward and cost. Recall that  the order of the actions does not matter as all selected DUs will be scheduled in the same transmission opportunity at time $\tau$  and none of the candidate DUs expires in the current transmission interval. For example, in Fig. \ref{fig:Full_trellis}, scheduling DU $F_1^K$ and then DU $F_3^{WZ}$ leads to the state $S_{2,2}$, which is the same state that can be reached by scheduling   DU $F_3^{WZ}$ first and DU $F_1^K$ afterwards. That state is reached with the same cost and reward in both cases.

Rule $1$ is equivalent to chosing first the key frames to be scheduled, before any other frame versions. It  reduces redundancy among branches without loss of optimality, but more importantly,  it  permits  to \emph{separate} reward-dependent DUs from reward-independent ones. 
We can then state  the second  rule.
\vspace{2mm}  \newline
{\bf \emph{Rule 2:}} If  the action $a$ corresponds to scheduling a WZ or P frame at state $S_{i,k}$, then  $a$ and all successor states/actions are replaced by   a single no action branch, leading to a final state $S^O_{i+1,k^{\prime}}$ with $C(S^O_{i+1,k^{\prime}})= C(S_{i,k})$, $\mathcal{A}(S^O_{i+1,k^{\prime}})=\mathcal{A}^p(S_{i,k})$ and with state value function $\widehat{V}_{\pi^{\star}}(S^O_{i+1,k^{\prime}})$, which corresponds to the optimal value function that can be reached by feasible scheduling of DUs in $\mathcal{A}^p(S_{i,k})$. 
\vspace{2mm}  \newline
Because of the separation of WZ/P sub-paths from key ones imposed by Rule 1, once a  WZ/P sub-path  starts, the optimal value function can be found by choosing the best set of reward-independent DUs in $\mathcal{A}^p(S_{i,k})$.   This problem can be written as follows: 
\vspace{2mm}

{\bf Problem 2:} 
\begin{quotation}
{\bf Init: } Let $\mathcal{A}^p(S_{i,k})$ be the set of candidate WZ or P DUs at state $S_{i,k}$.  The set of candidate DUs is defined as the acquired frames that do not expire within  the current transmission slot and that satisfy constraints provided in Eq.  \eqref{eq:PF_1}, Eq.  \eqref{eq:PF_3}, and  Eq. \eqref{eq:PF_4} in Problem 1. 
 Let $c_l$ and $\delta(a_l)$ be the transmission cost and reward, respectively, of DU $F_l\in \mathcal{A}^p(S_{i,k})$. Let $C(S_{i,k})$ be  the available BW. 
\\
{\bf Solve:} 
\begin{align} \label{eq:OPT-p2} 
\widehat{V}_{\pi^{\star}}(S_{i,k}): &\max_{\mathcal{T} \subseteq \mathcal{A}^p(S_{i,k})} \sum_{l \in  \mathcal{T}}  \delta(a_l)\\ \nonumber
&\text{s.t. } \sum_{l \in  \mathcal{T}} c_l\leq C(S_{i,k})
\end{align}
\end{quotation}
\vspace{2mm}

Problem 2 can be   solved by DP programming. It is actually as knapsack problem \cite{cormen2001} as shown in the following.    Let $\mathcal{A}^p_{1:j}(S_{i,k})\subset \mathcal{A}^p(S_{i,k})$ be the set of the first $j$ listed candidate DUs in $\mathcal{A}^p(S_{i,k})$. Let define $D[j,w]$ as $D[j,w] = \max_{\mathcal{T} \subseteq \mathcal{A}^p_{1:j}(S_{i,k})} \sum_{l \in  \mathcal{T}}  \delta(a_l)$ s.t.  $\sum_{l \in  \mathcal{T}} c_l\leq w$, where $c_l$ is the cost of the DU $a_l$. This means that $D[j,w]$ is the best cumulative reward obtained from selecting the best DUs among $\mathcal{A}^p_{1:j}(S_{i,k})$ whose transmission cost sums up to $w$. 
Since all DUs in $\mathcal{A}^p(S_{i,k})$ are   reward-independent, we can claim that 
\begin{align}\label{eq:DP}
D[j,w] = \max\{  D[j-1,w], D[j-1,w-c_j]  + \delta(a_j)\}.
 \end{align}  
 Thus, $D[|\mathcal{A}^p(S_{i,k})|,  C(S_{i,k})]$ is the solution to the Problem 2 and the   iterative equation \eqref{eq:DP}  allows to solve the optimization problem in Eq.  \eqref{eq:OPT-p2}   as dynamic programming problem (e.g., knapsack 0-1 problem) with a computational complexity of $O(|\mathcal{A}^p(S_{i,k})| C(S_{i,k}))$.

\begin{figure}[t]
 \begin{center}
\includegraphics[width=0.7\linewidth,   draft=false]{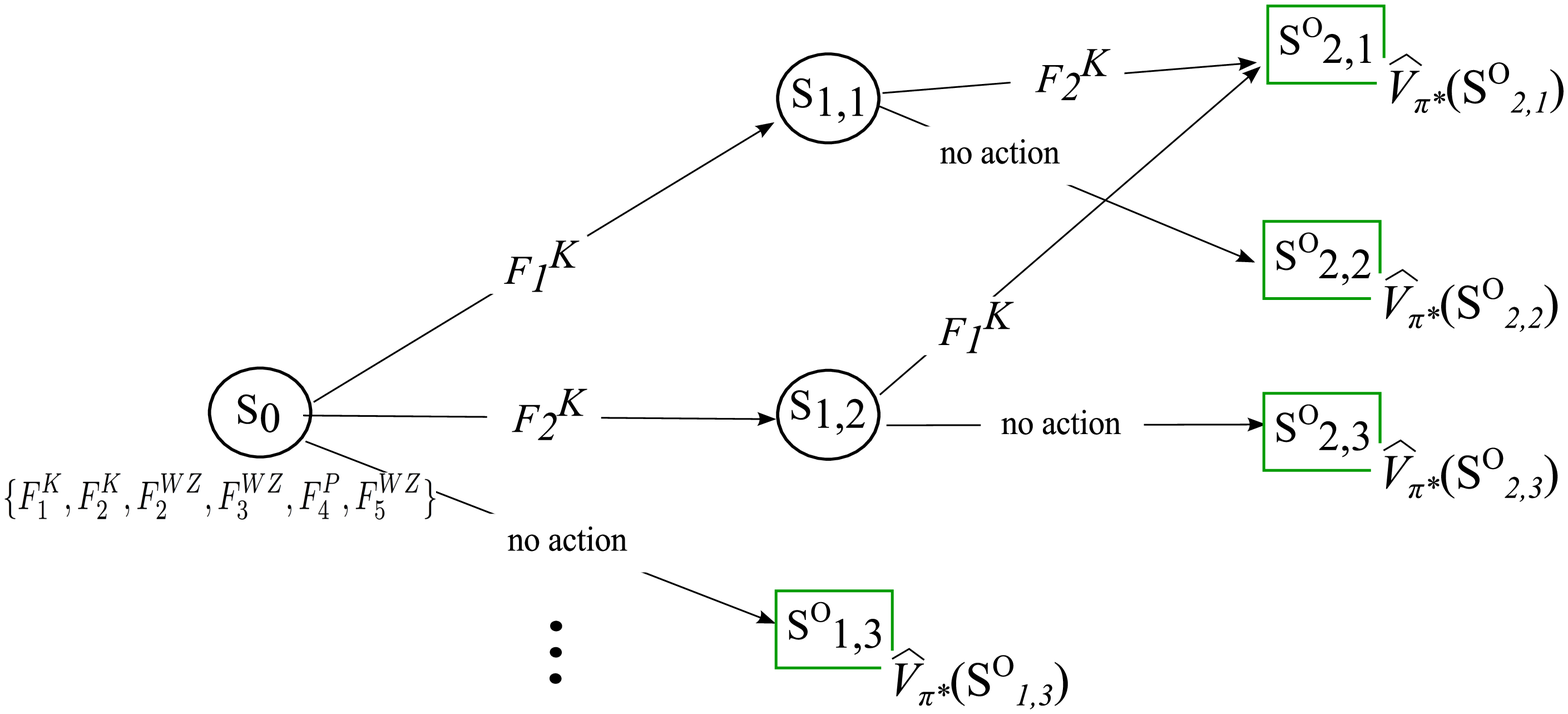}
\caption{ Equivalent trellis-based solution.  Only key frames can be   scheduled at   black nodes. A \emph{no action}    taken from state $S_{i,k}$ leads to a final state $S^O_{i+1,k^{\prime}}$.   At each final state,   \emph{WZ} or \emph{P}  can be scheduled following Problem 2.}   \label{fig:Optimal_solution_eq}
\end{center}
\end{figure}

\vspace{2mm}

 The trellis construction in Fig. \ref{fig:Full_trellis} can then be replaced by  the one in Fig. \ref{fig:Optimal_solution_eq}, where initial branches are constructed only for key actions and final states can be reached by taking no action from state $S_{i,k}$. In this case,  if $\mathcal{A}^p(S_{i,k})$ is not null and $C(S_{i,k})>0$, the successor final state  $S_{i+1, k^{\prime}}^O$ has a value function $\widehat{V}_{\pi^{\star}}(S_{i+1, k^{\prime}}^O)$ resulting from  Problem 2, which evaluates the best scheduling for  dependent DUs among the ones in $\mathcal{A}(S_{i+1, k^{\prime}}^O) = \mathcal{A}^p(S_{i,k})$.  The proof of optimality of our solving method with a modified trellis is provided  below.

\subsection{Proof of Optimality}
 Recalling that $\mathcal{A}(S_{i,k}) = \mathcal{A}^p(S_{i,k})\cup \mathcal{A}^k(S_{i,k})$, from 
 Eq. \eqref{eq:Bellman} we have  
\begin{align} \label{eq:decomp_Vs}
V_{\pi^{\star}}(S_{i,k}) &= \max_{a\in \mathcal{A}(S_{i,k})}\left\{ \delta(a) + \sum_{k^{\prime}} V_{\pi^{\star}}(S_{i+1,{k^{\prime}}}) P\left(S_{i+1,{k^{\prime}}}|S_{i,k}, a\right)  \right\} 
\nonumber \\
 &= \max\left\{  \max_{a\in \mathcal{A}^k(S_{i,k})}\left\{ \delta(a) + \sum_{k^{\prime}} V_{\pi^{\star}}(S_{i+1,{k^{\prime}}}) P\left(S_{i+1,{k^{\prime}}}|S_{i,k}, a\right)  \right\} , \right.\nonumber \\ 
 &\hspace{16mm} \left. \max_{a\in \mathcal{A}^p(S_{i,k})}\left\{ \delta(a) + \sum_{k^{\prime}} V_{\pi^{\star}}(S_{i+1,{k^{\prime}}}) P\left(S_{i+1,{k^{\prime}}}|S_{i,k}, a\right)  \right\}   \right\}  \nonumber \\
&= \max  \left\{ V_{\pi^{\star}}^k(S_{i,k}),  V_{\pi^{\star}}^p(S_{i,k}) \right\}
\end{align}
where $V_{\pi^{\star}}^k(S_{i,k})$ and $V_{\pi^{\star}}^p(S_{i,k})$ represent the value function of state $S_{i,k}$ under the best policy $\pi^{\star}$, characterized by the scheduling of only key frames and WZ/P frames, respectively. 
The decomposition allows to distinguish the best action taken at state $S_{i,k}$ as a key $\left(\max  \left\{ V_{\pi^{\star}}^k(S_{i,k}),  V_{\pi^{\star}}^p(S_{i,k}) \right\}\right.$ = $\left.V_{\pi^{\star}}^k(S_{i,k})\right)$  or  a WZ or P frame $\left(\max  \left\{ V_{\pi^{\star}}^k(S_{i,k}),  V_{\pi^{\star}}^p(S_{i,k}) \right\} = V_{\pi^{\star}}^p(S_{i,k})\right)$.
The state value function $V_{\pi^{\star}}^k(S_{i,k})$ assumes that a key frame is scheduled from state $S_{i,k}$, thus Rule 2 does not apply to this set of possible actions. On the contrary, $V_{\pi^{\star}}^p(S_{i,k})$ is the state value function under the policy of scheduling a WZ or P frames in state $S_{i,k}$ and in all future states, from Rule 1.  
We  now focus on $V^p(S_{i,k})$ and expand it  as follows
\begin{align}
 V_{\pi^{\star}}^p(S_{i,k})  &=
 \max_{a\in \mathcal{A}^p(S_{i,k})}\left\{ \delta(a) + \sum_{k^{\prime}}  P\left(S_{i+1,{k^{\prime}}}|S_{i,k}, a\right) V_{\pi^{\star}}(S_{i+1,{k^{\prime}}}) \right\}  \nonumber \\
 &= \max_{a\in \mathcal{A}^p(S_{i,k})}\left\{ \delta(a) + \sum_{k^{\prime}}P\left(S_{i+1,{k^{\prime}}}|S_{i,k}, a\right) \right. \nonumber\\
& \hspace{5mm} \left. \left( \max_{a^{\prime}\in \mathcal{A}(S_{i+1,j})}\left\{ \delta(a^{\prime}) + \sum_{k^{\prime\prime}}P\left(S_{i+2,{k^{\prime\prime}}}|S_{i+1,k^{\prime}}, a^{\prime}\right) V_{\pi^{\star}}(S_{i+2,m})  \right\} \right)
  \right\} 
 \end{align}
Noting that $\mathcal{A}(S_{i+1,j}) =  \mathcal{A}^p(S_{i+1,j})$ because of Rule 1, the above expression is equivalent to  
 \begin{align}
 \label{eq:v_opt}
 V_{\pi^{\star}}^p(S_{i,k})  &= \max_{\substack{
a  \in \mathcal{A}^p(S_{i,k}), \\ a^{\prime} \in \mathcal{A}^p(S_{i+1,j})}
}
\left\{ \delta(a) + \delta(a^{\prime}) + \sum_{k^{\prime\prime}}P\left(S_{i+2,{k^{\prime\prime}}}|S_{i,k}, a, a^{\prime}\right) V_{\pi^{\star}}(S_{i+2,m}) \right\} 
\nonumber \\
  &= \max_{ \substack{
a,  a^{\prime}  \in \mathcal{A}^p(S_{i,k}), a \neq a^{\prime}} }
\left\{ \delta(a) + \delta(a^{\prime}) + \sum_{k^{\prime\prime}}P\left(S_{i+2,{k^{\prime\prime}}}|S_{i,k}, a, a^{\prime}\right) V_{\pi^{\star}}(S_{i+2,m}) \right\} 
\end{align}
where    we have   used the property that    $\mathcal{A}^p(S_{i+1,j}) =  \mathcal{A}^p(S_{i,k}) \setminus a$ because of the coding independency among P and WZ DUs. 
Denoting by $I$ the maximum number of actions that can be taken from $S_{i,k}$ under the best policy $\pi^{\star}$ and expanding Eq. \eqref{eq:v_opt} till the final state we get  
\begin{align}\label{eq:opt}
V_{\pi^{\star}}^p(S_{i,k})    &= 
\max_{  \mat{a}   \in \mathcal{A}^p(S_{i,k})}
\left\{\sum_{q=1}^I \delta(a_{q})  + \sum_{k^{\prime}}P\left(S_{i+I,{k^{\prime}}}|S_{i,k},  \mat{a}   \right) V_{\pi^{\star}}(S_{i+I,k^{\prime}}) \right\} \nonumber
\\  
  &=
\max_{  \mat{a}   \in \mathcal{A}^p(S_{i,k})}
\left\{\sum_{q=1}^I \delta(a_{q})  \right\}
\end{align}
where $\mat{a} =[a_1, a_{2}, \ldots a_{I}]$ is the action vector and  the last equality holds since all final states in the original trellis are set to $0$. 
This means that  $V_{\pi^{\star}}^p(S_{i,k})$ corresponds to the gain achieved by solving  the optimization problem in Eq. \eqref{eq:OPT-p2}, namely $V_{\pi^{\star}}^p(S_{i,k})=  \widehat{V}_{\pi^{\star}}(S_{i+1,j}^O)$, with $S_{i+1,j}^O$ being the state that can be reached from $S_{i,k}$ by taking no actions. 
Then, Eq.  \eqref{eq:decomp_Vs} is equivalent to
\begin{align} 
V_{\pi^{\star}}(S_{i,k}) &= \max  \left\{ V_{\pi^{\star}}^k(S_{i,k}),   \widehat{V}_{\pi^{\star}}(S_{i+1,j}^O)) \right\}
\end{align}
This proves that Rule 2 permits to reach optimality. 
$\square$

\vspace{2mm}

We have described above a novel   trellis-based solution to optimize the problem in Eq. \eqref{eq:PF}. To simplify the computational complexity of the solution, that would be otherwise exponential, we have proposed two scheduling rules. These allow to   decouple the   actions of scheduling key (reward-dependent) frames   from WZ/P (reward-independent) frames. From this novel decomposition, we can then reduce any  WZ/P sub-path  in the trellis to an equivalent  final state whose state value function is the solution of a simple knapsack $0-1$ optimization problem.

\section{Simulation Results}
\label{sec:results}

\subsection{Simulation Setup}

We provide now simulation results for a multi-camera scenario where data have to be sent to a central server over a bottleneck channel.   We start the scheduling optimization at $\tau=1$ and  set the following transmission opportunities  every $\Delta t$. Each transmission opportunity is characterized by a channel rate $C_{\tau}$.    At this new scheduling opportunity, a new optimization is performed over the successive   time slot. We proceed similarly till the end of the simulation, which in our case corresponds to the expiration time   of the last frame of the video sequence.  
 
Our simulations are carried out with the ``Ballet"   video sequence \cite{web_microsoft_ballet_break}, which consists of $N_f = 100$ frames, at a resolution of $S_R = 768 \times 1024$ pixel/frame and $F_R=15$ frames per second. The total number of camera is $8$. Since  ``Ballet" is a quite static video sequence where the spatial correlation model does not substantially change over time and the temporal correlation is extremely large,  we also created a synthetic $16$-views sequence with a more dynamic correlation model  to test our algorithm over a more challenging scenario. In this synthetic sequence,   the spatial correlation model substantially changes every $20$ frames. In practice this corresponds to  a moving obstacle in the scene, or to moving cameras.    For both sequences, we study the performance of our algorithms in different configurations, for different camera setups, different  users' behavior and for different dynamics of the channel bandwidth.

The image correlation used in decoding and reconstruction of the different frames is characterized by two parameters, namely $\rho_\text{S}$ and $\rho_\text{T}$. We   denote by $\rho_\text{S}$ the number of  spatially  correlated cameras and we  assume that each view is correlated to at most $\rho_\text{S}/2$ neighbor 
views,   if available, on both the left and the right sides.   
The correlation in time is denoted by $\rho_\text{T}$, which corresponds to the number of correlated images in the same camera view.  Both $\rho_\text{T}$ and $\rho_\text{S}$ represent the \textit{maximum} number of correlated images in the time and space domain, respectively. The control parameters $\rho_\text{T}$  and $\rho_\text{S}$  take different values  in our simulations in order to study the behavior of the scheduler for    different neighborhood, as defined in Eq. \eqref{eq:neighborhood}.   Then, the \textit{actual} level of correlation $\rho$ experienced in each single frame depends   on the video content.  It is computed as the portion of image that can be reconstructed by each image in the neighborhood. 
We refer the reader to \cite{tech_report:12} for further details on the construction of the correlation values. 

The network scenarios  considered in our simulations are characterized by either \emph{static or dynamic channels}. The former means that the channel bandwidth is constant over the entire streaming session, while the latter consider a dynamic behavior of the channel. In this case, we model the channel as a 2-state Markov model where bad and good states identifies two different  values for the available channel bandwidth.  We denote by  $p$ the transition probability, i.e., the probability of change state,  in the Markov model. For each video sequence a realization of the dynamic channel is considered and the scheduling performance is evaluated for that specific channel realization. This is iterated for $100$ loops to compute average performance with   channel dynamics.
We further study two models for user interactivity, namely   \emph{static or dynamic multiview navigation}. 
In the case of static navigation, we assume that the view transition probability $w_{jl}=0, $ for $j\neq l$ and $w_{jj}=1$. We also consider a uniform camera popularity, i.e., $P_l=1/M$, with $M$ being the number of camera views. This scenario emulates a static scene  where  there is no a peak of interest in specific view and no interest in changing  viewpoints. On the other hand,   dynamic navigation is the scenario in which    the navigation path evolves over time, to follow changes in the scene or change of preferences for users.  From a given frame $F_{t,m}$ the user can navigate to neighboring views with probability $w_{m,l}$. In particular, users most likely select views more on the right (left) if the scene is moving to the right (left). As a result, the camera popularity for the first acquired frames is $1/M$, while for all successive instants the popularity is derived from the transition probabilities, i.e.,  $P_{t,m}= \sum_{l} P_{t-1,l} w_{lm}$.

The performance results are given by the average quality, computed as  PSNR averaged over the views,  with the average weighted by the camera popularity\footnote{The camera popularity evolves over time for  dynamic navigation paths, while it is constant   for static navigation paths.}. This leads to an average PSNR value for each acquisition time. Alternatively,  we also provide the popularity-weighted PSNR values averaged both in time and in space. In   case of dynamic channel settings,  the latter metric is also averaged over the $100$ simulated loops, while  the PSNR over time is provided for a representative realization rather than the behavior averaged over the loops. This allows to better observe the quality oscillations experienced by users.   Note that, even if some frames are decoded at high quality, the average  PSNR of the reconstructed scene might be in the low PSNR range in challenging transmission conditions.   
 
 Finally, we compare the proposed algorithm  to three baseline algorithms: two scheduling strategies (``BL, Cont=0" and ``BL, Cont=1") for a pre-selected coding and camera selection strategy,    our previous scheduling solution (``Toni et al."~\cite{Toni:J14}) where a simplistic coding is considered and no dynamic navigation path is taken into account, and the well known ``RaDiO" algorithm~\cite{ChoMia:J06}.  In particular, ``BL, Cont=0" considers an a priori camera selection  and a coding strategy optimized  based on the spatial correlation that exists between views at the beginning of the sequence. This means that we consider a pre-selected coding structure and camera priority order;  at every transmission opportunity,  we schedule the sufficient number of DUs to reach the channel bandwith. In practice, we have considered the camera selection algorithm  in \cite{WanDaiAky:C11} and   we have extended it to a  coding and camera selection algorithm  such that we can have  a fair comparison with our algorithm.   The second baseline method, ``BL, Cont=1", is an improved version of the previous one, where we assume that the coding and camera selection is updated at every acquired frame. This means that the selection constantly considers an updated and correct correlation model, but it neglects the channel information in the optimization of the packet scheduling. Finally, the packet scheduling optimization ``Toni et al."  uses a correlation-aware packet scheduling optimization that is refined at every transmission; the camera popularity is considered in the optimization but there is no consideration of the navigation path and quality variations, and only key frames are used as candidate DUs. 
The last baseline algorithm that we have implemented is the  ``RaDiO" one, whose scheduling optimization   has been extended to multiview streaming. We have considered that   each frame candidate for being scheduled is a DU. Each DU has its own policy vector (deciding if sending the DU and in which encoded version) and the optimal scheduling strategy is evaluated iterating the optimization over each considered DU, following the same procedure as in~\cite{ChoMia:J06}.

 In the following,   the PSNR of the reconstructed scene  is first evaluated from the rate-distortion model  described in Sec.~\ref{sec:RD}. Later, we validate our findings by experiments with actual reconstruction of the video frames at the decoder. 
 
\subsection{Average distortion minimization}

 \begin{figure}[t]
\subfigure[Synthetic Sequence,  $T_A=4$, $T_D=1$,  $C=\{2,1\}$]{
\includegraphics[width=0.46\linewidth,  draft=false]{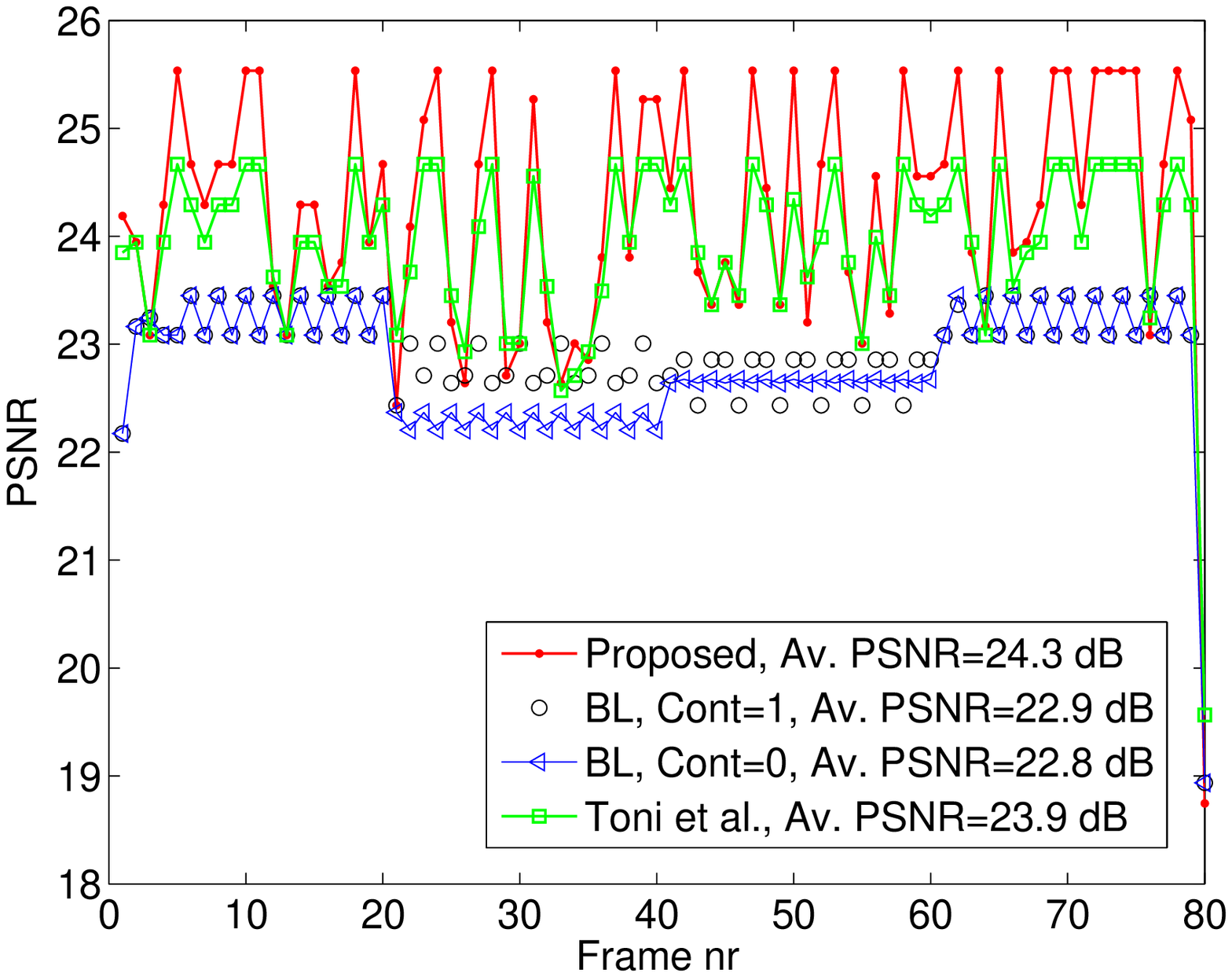} }  \hfill   
\subfigure[Ballet Sequence,  $T_A=1$, $T_D=2$,   $C=\{1.5,1\}$]{
\includegraphics[width=0.46\linewidth,  draft=false]{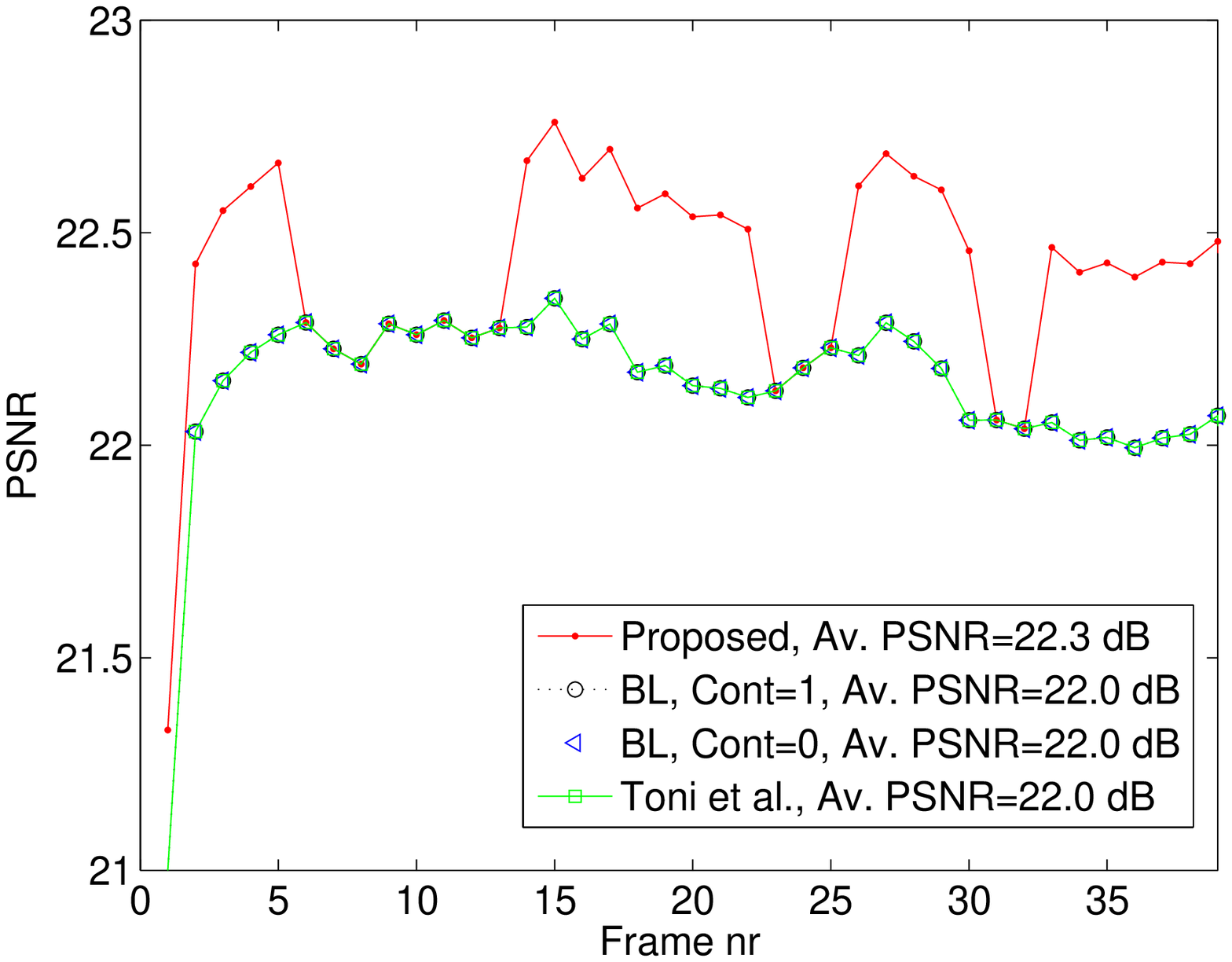} } 
\caption{Temporal PSNR evolution for different scheduling algorithms ($\rho_S=4, \rho_T=1$, static navigation path and \emph{dynamic channel}, $p=0.8$).  } \label{fig:ViewTime_weighted_Ts1_Ta1_Td2_SQ1_P8_Lambda0_SNav1_Tc1_Ts4_Nview8_BW1_5__1}
\end{figure}

\begin{figure}
\begin{center}
\includegraphics[width=0.46\linewidth,  draft=false]{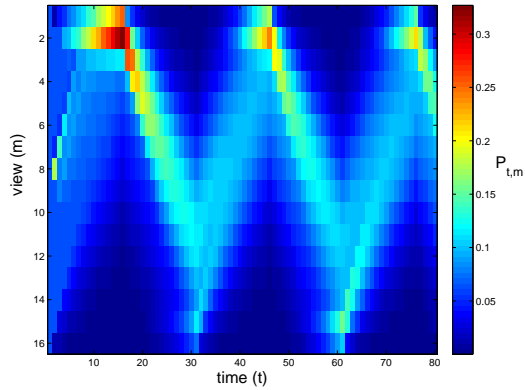}   
\caption{View popularity $P_{t,m}$ for the Synthetic video sequence with  dynamic navigation.} \label{fig:camera_popularity}
\end{center}
\end{figure}

We first look at the behavior of the scheduling strategies in the case of dynamic channels when the objective function does not consider quality variations, i.e., $\lambda=0$ in Eq. \eqref{eq:PF}.   For the sake of clarity, we first compare our scheduling algorithm with ``BL" and ``Toni et al." baseline algorithms. Then, we   provide a comparison with the ``RaDiO" method.  
In Fig. \ref{fig:ViewTime_weighted_Ts1_Ta1_Td2_SQ1_P8_Lambda0_SNav1_Tc1_Ts4_Nview8_BW1_5__1}, we depict the popularity-weighted PSNR (averaged over the views) as a function of the   frame index for both Synthetic and Ballet sequence. The navigation path is static but the channel is dynamic, with $p=0.8$. For the Synthetic sequence,  we have the channel states defined as $C=\{2,1\}$,  which means that the available bandwidth is two times (one time) the transmission cost of a key frame  in good (bad) channel conditions,    while for the Ballet sequences the channel states are $C=\{1.5,1\}$. The results are averaged over several simluations, each one considering a specific realization of the channel. For each realization, all algorithms are tested in order to have a fair comparison among them. 
For both video sequences,  the   variations of the channel leads to a substantially varying PSNR over time. This is one of the main motivation for taking into account the variations of the quality in the objective function (i.e., $\lambda\neq 0$) as shown in the following subsection. Despite these variations, we still have that the proposed algorithm outperforms baseline algorithms in most of the time slots,  as it can be observed from the average PSNR values. We can also observe that,   for the Synthetic sequence,  the   gain is larger than the gain achieved by the Ballet sequence. This is mainly due to the fact that the Ballet sequence is highly correlated both in time and space and the correlation is very uniform in both dimensions. This makes the streaming scenario less challenging. Hence, there is less room for improvement by our algorithm. On the contrary, the Synthetic sequence has many obstacles in the scene, thus non-optimal scheduling   substantially affects the experienced quality.

 \begin{figure}
 \begin{center}
\subfigure[Synthetic Sequence,  $T_A=4$, $T_D=1$, $C=2$]{
\includegraphics[width=0.46\linewidth,  draft=false]{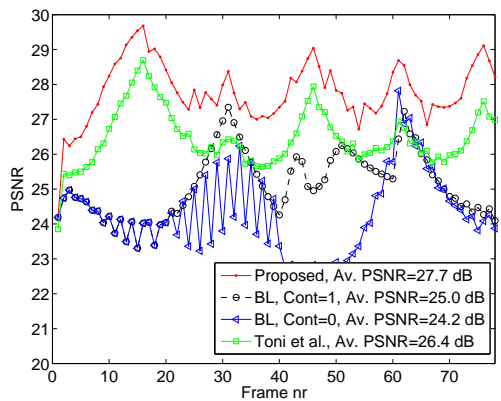} }  \hfill   
\subfigure[Ballet Sequence,  $T_A=1$, $T_D=3$, $C=1.5$]{
\includegraphics[width=0.46\linewidth,  draft=false]{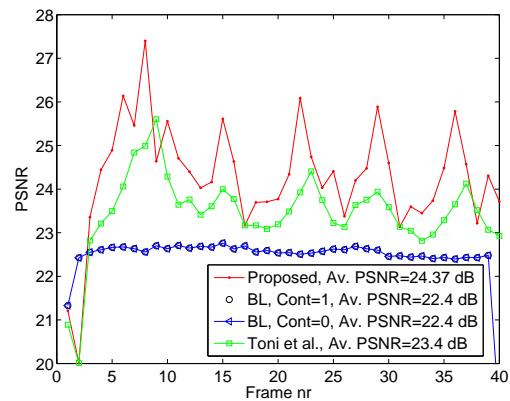} }   \caption{Temporal PSNR evolution for different scheduling algorithms ($\rho_S=4$,  $\rho_T=1$,  \emph{static channel and   dynamic  navigation path}).  } \label{fig:ViewTime_weighted_Ts1_Ta4_Td1_SQ1_P0_Lambda0_SNav0_Nview16}
\end{center}
\end{figure}

Finally,   we also study the performance of different algorithms in a scenario in which the channel is static while the navigation path is dynamic. In Fig. \ref{fig:camera_popularity}, we depict the simulated  frame popularity resulting from a dynamic navigation path. It simulates a scenario in which the subject of interest constantly move from left to right and back. The same type of navigation is used for both sequences. In Fig. \ref{fig:ViewTime_weighted_Ts1_Ta4_Td1_SQ1_P0_Lambda0_SNav0_Nview16}, the mean PSNR (popularity-weighted average  over views) is provided as function of the frame index for both sequences and for $\rho_S=4, \rho_T=1$.  In both cases,  we observe the gain obtained by our algorithm  that  constantly updates the optimal scheduling to the dynamic navigation path. 
This is deduced by comparing the proposed algorithm and the ``Toni et al.", which also refines the scheduling policy at each transmission opportunity,  with the BL algorithms, which have a static scheduling optimization.   Since   the algorithm  ``Toni et al."  also tracks camera popularity variations, it is able to perform quite well in the considered scenario, but still it suffers from a simplistic coding scheme.

\begin{figure}
 \begin{center}
\subfigure[$\rho_T=0$]{
\includegraphics[width=0.46\linewidth,  draft=false]{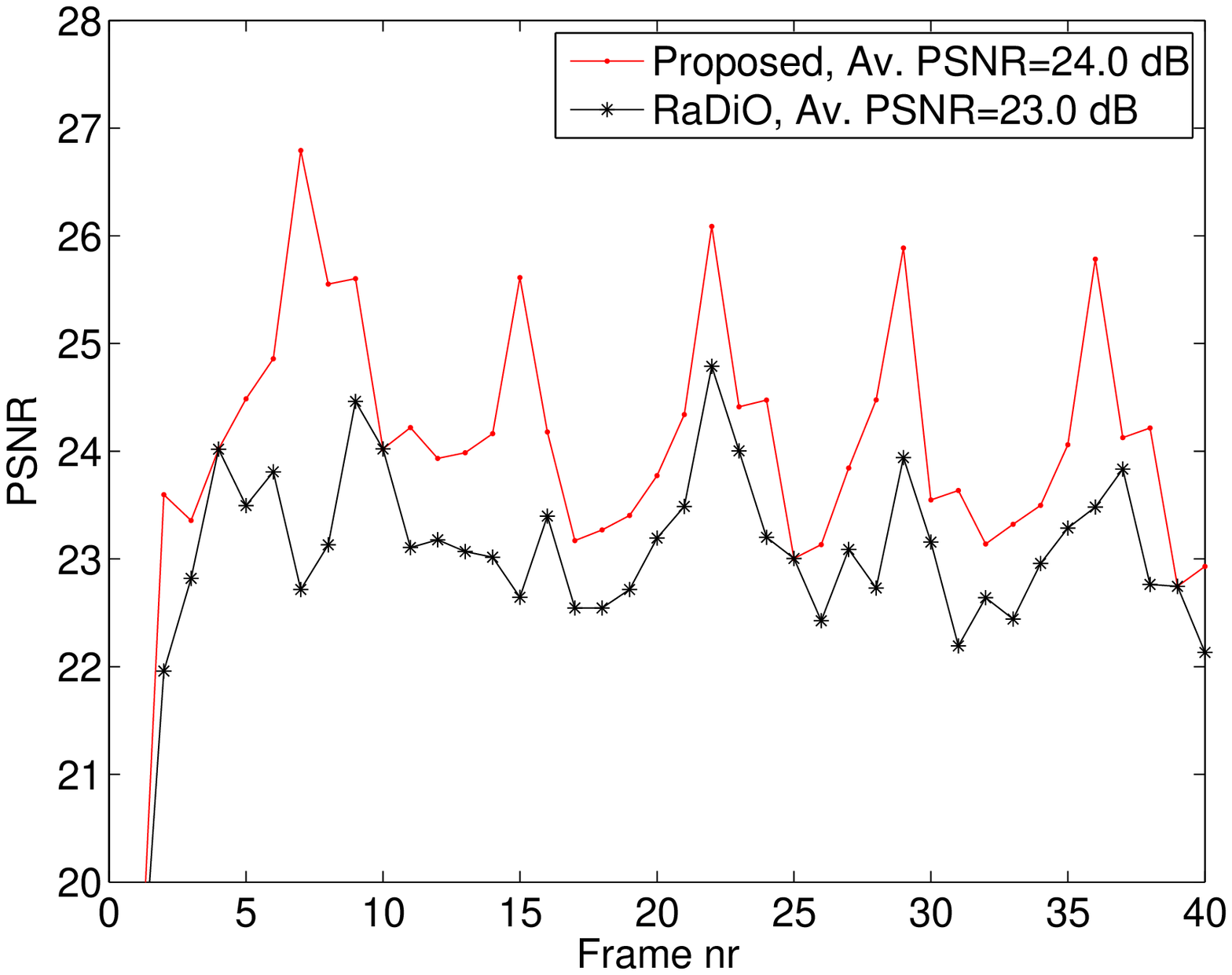} }  \hfill   
\subfigure[$\rho_T=1$]{
\includegraphics[width=0.46\linewidth,  draft=false]{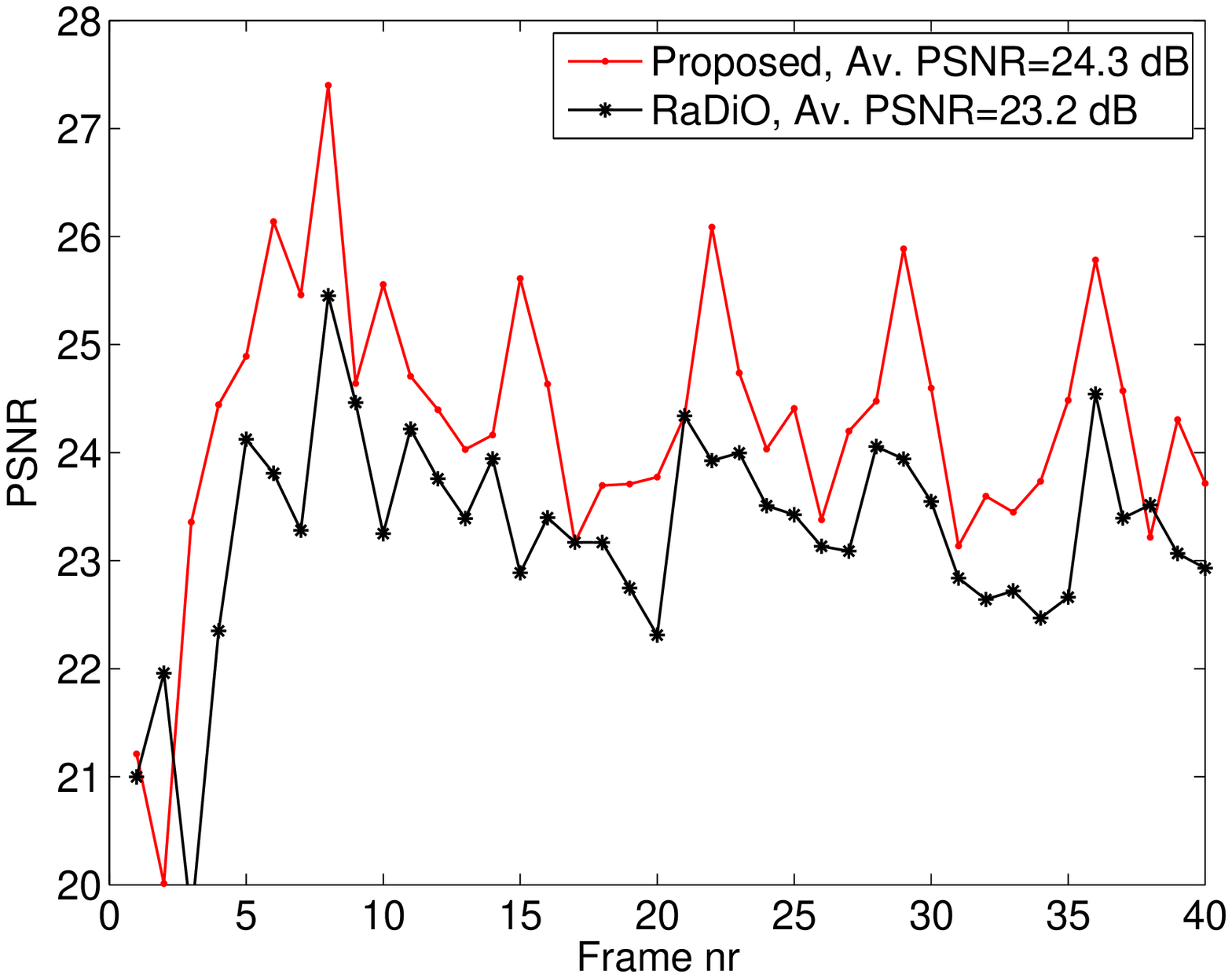} }   \caption{Temporal PSNR evolution for  the proposed algorithm and the RaDIO one for the Ballet Sequence ($T_A=1$, $T_D=3$, $C=1.5$,   $\rho_S=4$,   \emph{static channel and   dynamic  navigation path}).  } \label{fig:RADIO}
\end{center}
\end{figure}

For the sake of completeness  we also provide a 
comparison with the ``RaDIO" algorithm for the Ballet sequence with the following settings: $T_A=1$, $T_D=3$, $C=1.5$, and    $\rho_S=4$ (see Fig. \ref{fig:RADIO}). A static channel and a  dynamic navigation with the same model as above are considered.   We also simulated other settings and we obtained similar results. Thus, for brevity here we only provide one simulated setting.   The results are shown for two different levels of   temporal correlation $\rho_T$. Due to the iterative solving method, the  ``RaDIO" method has a reduced complexity, but does guarantee optimality~\cite{ChoMia:J06}. This leads to a loss of performance with respect  to the algorithm proposed in this paper that   reaches the optimal scheduling policy.

\begin{figure}
\subfigure[Average PSNR]{
\includegraphics[width=0.46\linewidth,  draft=false]{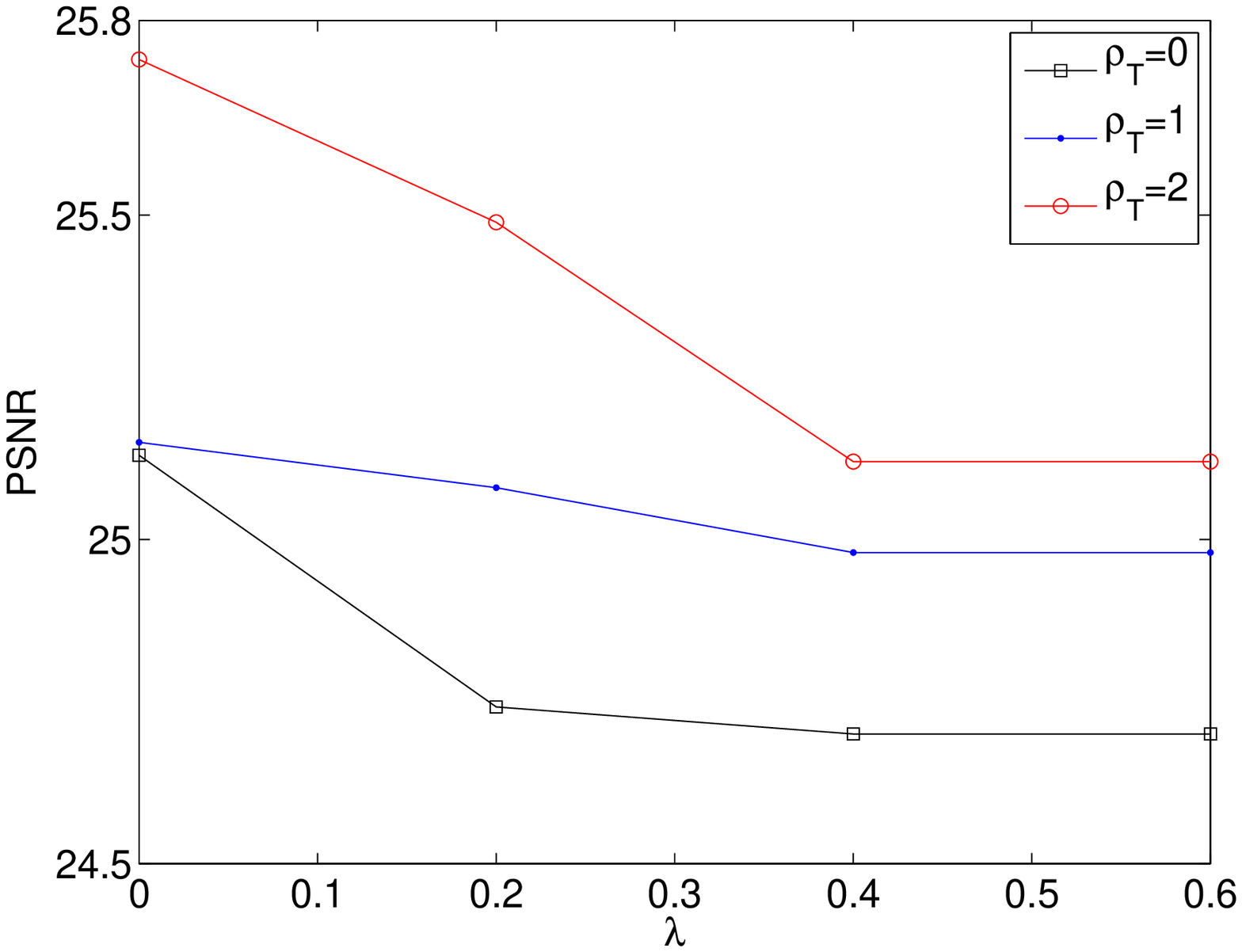} }  \hfill   
\subfigure[Average Variance]{
\includegraphics[width=0.46\linewidth,  draft=false]{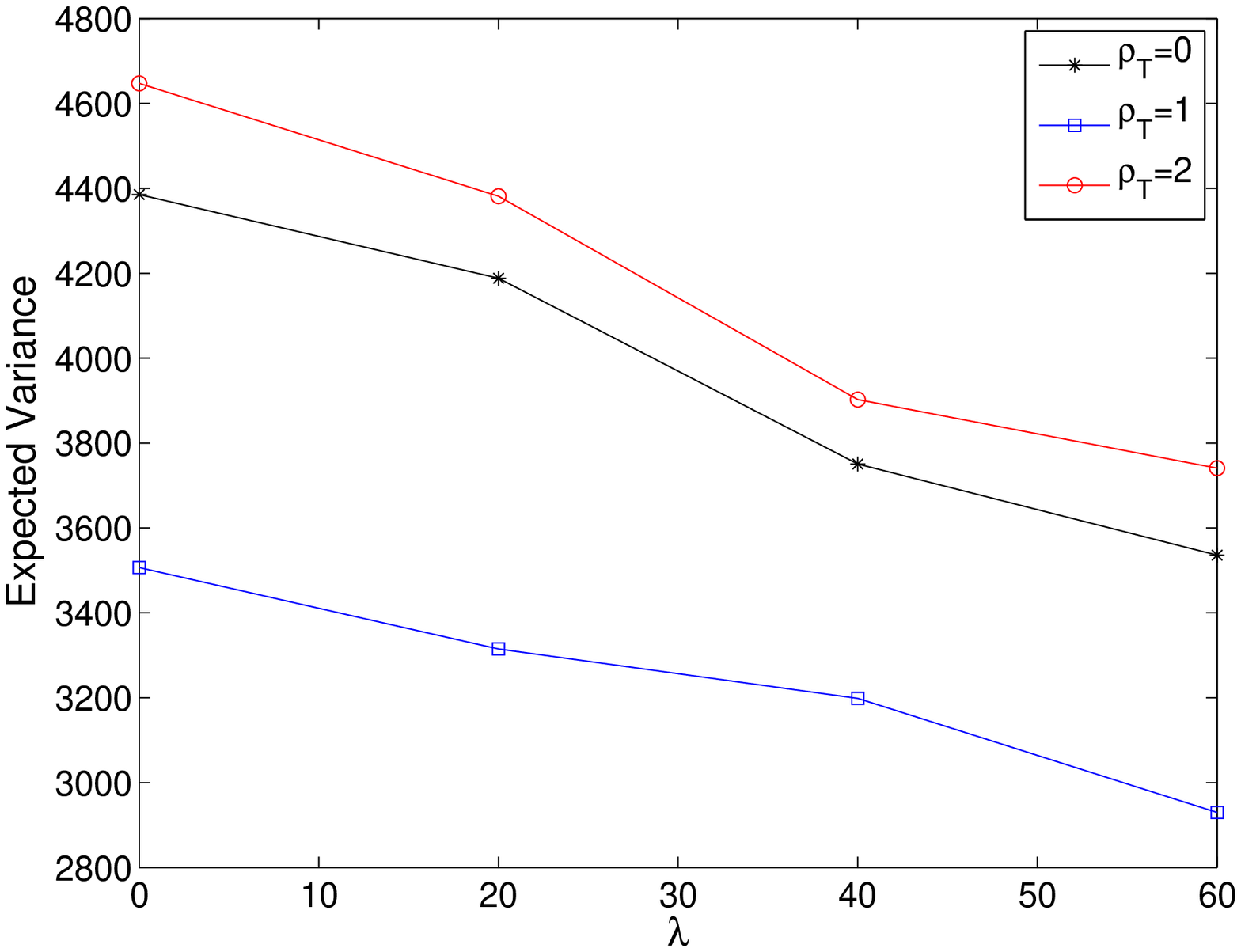} }   \caption{Average PSNR  and quality variance vs. optimization parameter $\lambda$ for Synthetic sequence for our scheduling algorithm ($T_A=4$, $T_D=1$,  $\rho_S=2$, static channel and dynamic navigation path). } \label{fig:ViewTimeConcealment_Ts1_Ta4_Td1_SQ1_P0_SNav1_Ts2_Nview16}
\end{figure}

With the above results, we have shown that the proposed algorithm outperforms competitor scheduling ones, but  it still suffers of large quality variations over time. In the following, we study the effect of  including quality variations  in the objective function for the proposed algorithm.   Baseline algorithms are not investigated in the following. As shown above, even when the proposed algorithm aims at minimizing only the weighted distortion, the baseline algorithms cannot compete with our solution. The main reason is that no information about users' interactivity is considered. Thus, we do not expect 
these algorithms to be able to compete with our solution when the objective function further includes  the quality variations over the navigation paths.    

\subsection{Quality variations minimization}

 \begin{figure}
\begin{center}
 \subfigure[Navigation path   starting from view $4$]{
\includegraphics[width=0.47\linewidth,  draft=false]{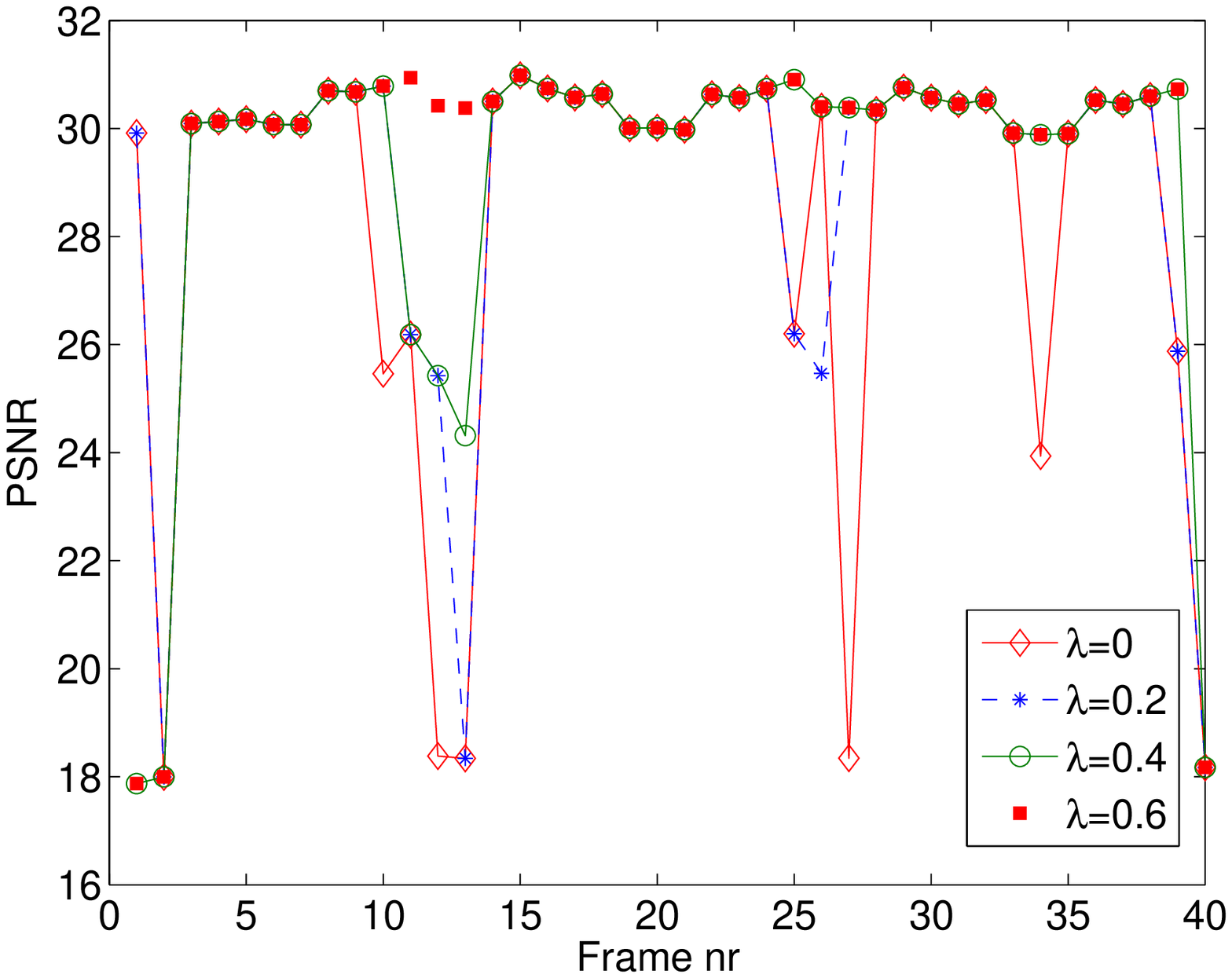}  }
\subfigure[Navigation path   starting from view $6$]{
 \includegraphics[width=0.47\linewidth,  draft=false]{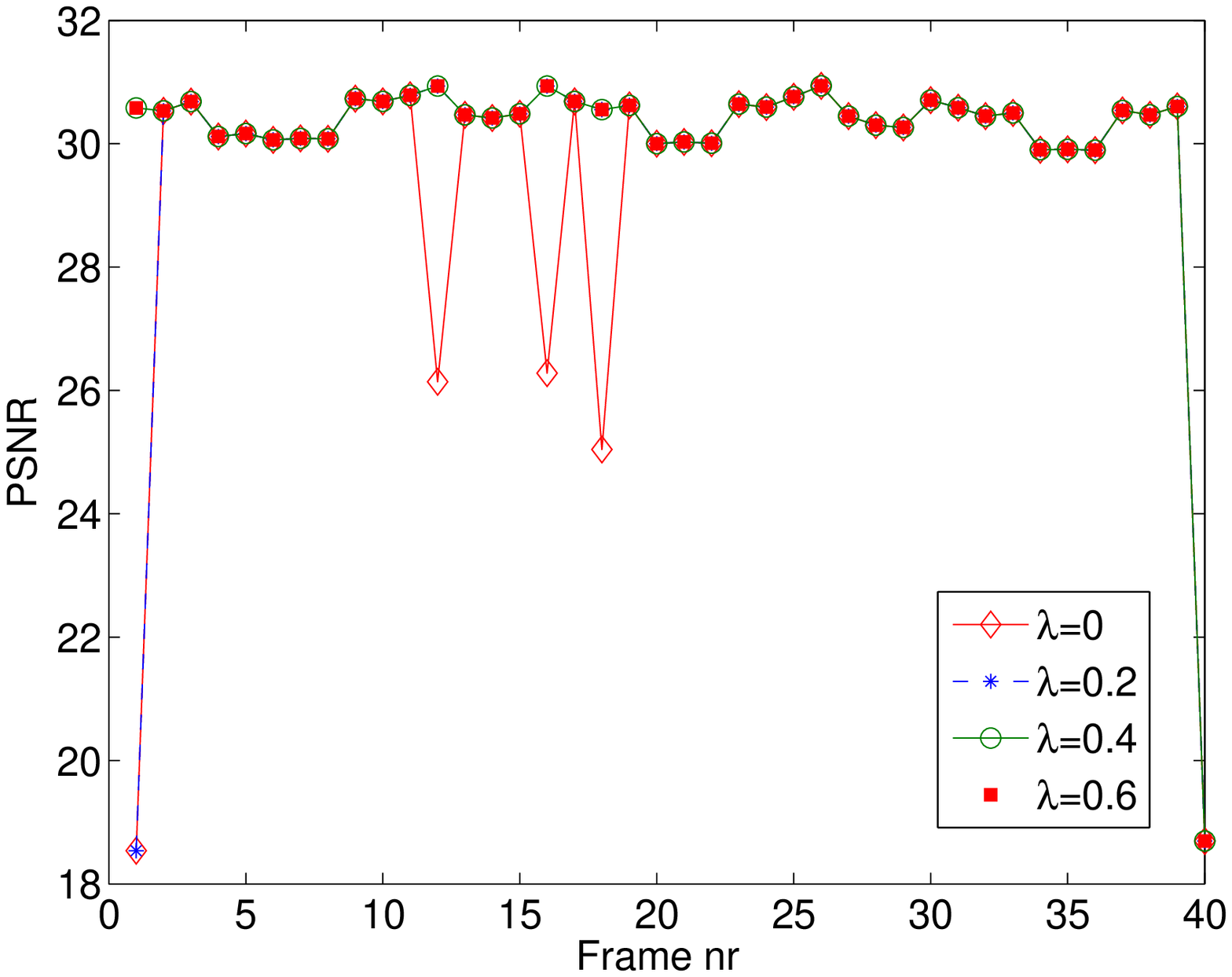} }
 \caption{Average PSNR as function of the frame index over the most likely navigation path for Ballet  sequence ($\rho_T=0, \rho_S=2$, $T_A=1$, $T_D=3$, $C=1.5$, static channel and dynamic navigation path). } \label{fig:MLP_Ts1_Ta1_Td3_SQ1_P0_ComparisonLambda_SNav0_Tc0_Ts2_Nview8}
\end{center}
\end{figure}

We are now interested in the behavior of the optimal scheduling policy when the objective function minimizes both the expected distortion and the expected variations of the quality   over the  navigation paths. Thus, in the following we study the performance of schedulers of both the average  (popularity-weighted) quality and the variance of the quality.    The variation of the quality is evaluated as in Eq.  \eqref{eq:PF}, which computes   a popularity-weighted variance.  

Fig. \ref{fig:ViewTimeConcealment_Ts1_Ta4_Td1_SQ1_P0_SNav1_Ts2_Nview16} depicts both expected quality and variance as function of  the optimization parameter $\lambda$, which trades off average quality and quality variations in the objective function of Eq. \eqref{eq:PF},  for $T_A=4$, $T_D=1$, $\rho_S=2$, and different levels of temporal correlation for the Synthetic sequence. A static channel and a dynamic navigation according to the model of Fig. \ref{fig:camera_popularity} are used in these simulations. As expected, the larger   $\lambda$, the more the variance becomes crucial in the optimization; the quality variations get smaller at the price of a reduced average quality. 
Similar trends can be observed for the Ballet sequence. 

\begin{figure} 
\subfigure[Navigation path starting from view  $1$]{
\includegraphics[width=0.45\linewidth,  draft=false]{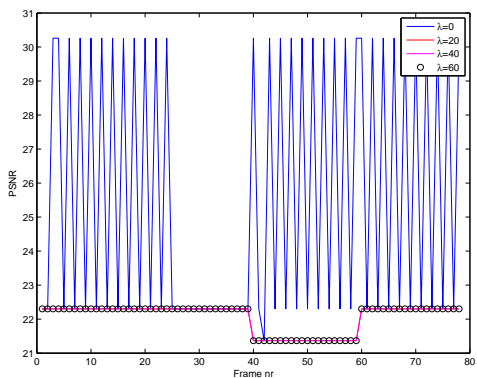} }  \hfill   
\subfigure[Navigation path starting from view  $2$]{
\includegraphics[width=0.45\linewidth,  draft=false]{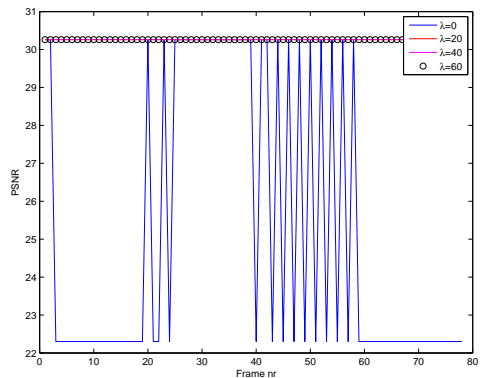} } \hfill
\subfigure[Navigation path starting from view  $4$]{
\includegraphics[width=0.45\linewidth,  draft=false]{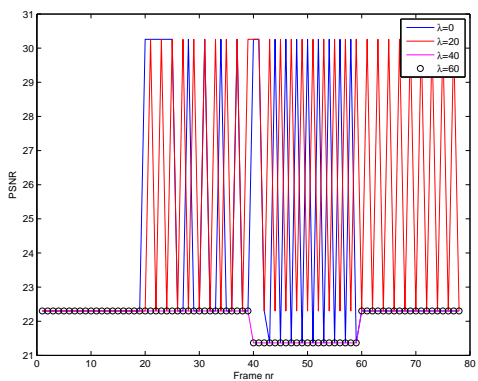} }\hfill
\subfigure[Navigation path starting from view  $8$]{
\includegraphics[width=0.45\linewidth,  draft=false]{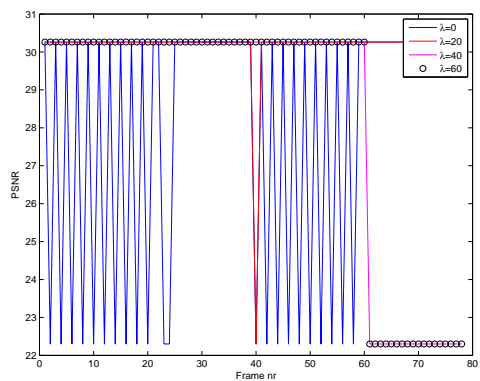} } \hfill
  \caption{PSNR vs. frame number for Synthetic sequence in the setting of $T_s=1$,  $T_A=4$, $T_D=1$, $\rho_S=2, \rho_T=1$, static channel, and dynamic navigation path. } \label{fig:MLP_Synth_Dyn_BL0Cont_1_Ts1_Ta4_Td1_SQ1_P0_SNav1_rhoT1}
\end{figure}

To give a better understanding about the impact of   a reduced variance,  we have evaluated the temporal evolution of the quality   over the most likely navigation path that starts from view $4$ or view $6$ (see Fig. 
\ref{fig:MLP_Ts1_Ta1_Td3_SQ1_P0_ComparisonLambda_SNav0_Tc0_Ts2_Nview8}).  It is worth noting that the quality perceived with $\lambda=0$ is subject to important fluctuations over time. The larger 
$\lambda$, the less these fluctuations  till the case of $\lambda=0.6$, where the quality variations are the smallest in these simulations. It is worth noting that limiting the variations might result in keeping the average   quality constant at a low value. 
 However, this is still expected to lead  to a quality of experience that   is better than a highly varying image quality. 
The case of Synthetic sequence is provided in  Fig. \ref{fig:MLP_Synth_Dyn_BL0Cont_1_Ts1_Ta4_Td1_SQ1_P0_SNav1_rhoT1} in the setting of $T_s=1$,  $T_A=4$, $T_D=1$, $\rho_S=2, \rho_T=1$, static channel, and dynamic navigation path.  In the figure we show the quality over the most likely navigation path when starting from different views. 
 It can be observed that reducing the quality variations experienced over the navigation path does not always lead to a large quality. Starting from View $1$ and View $4$, the most likely path will be forced to remain at a low-quality level but constant, allowing other paths to be constant at high quality level. 

\begin{figure}
\subfigure[Uniform and non-uniform navigation paths]{
\includegraphics[width=0.46\linewidth,  draft=false]{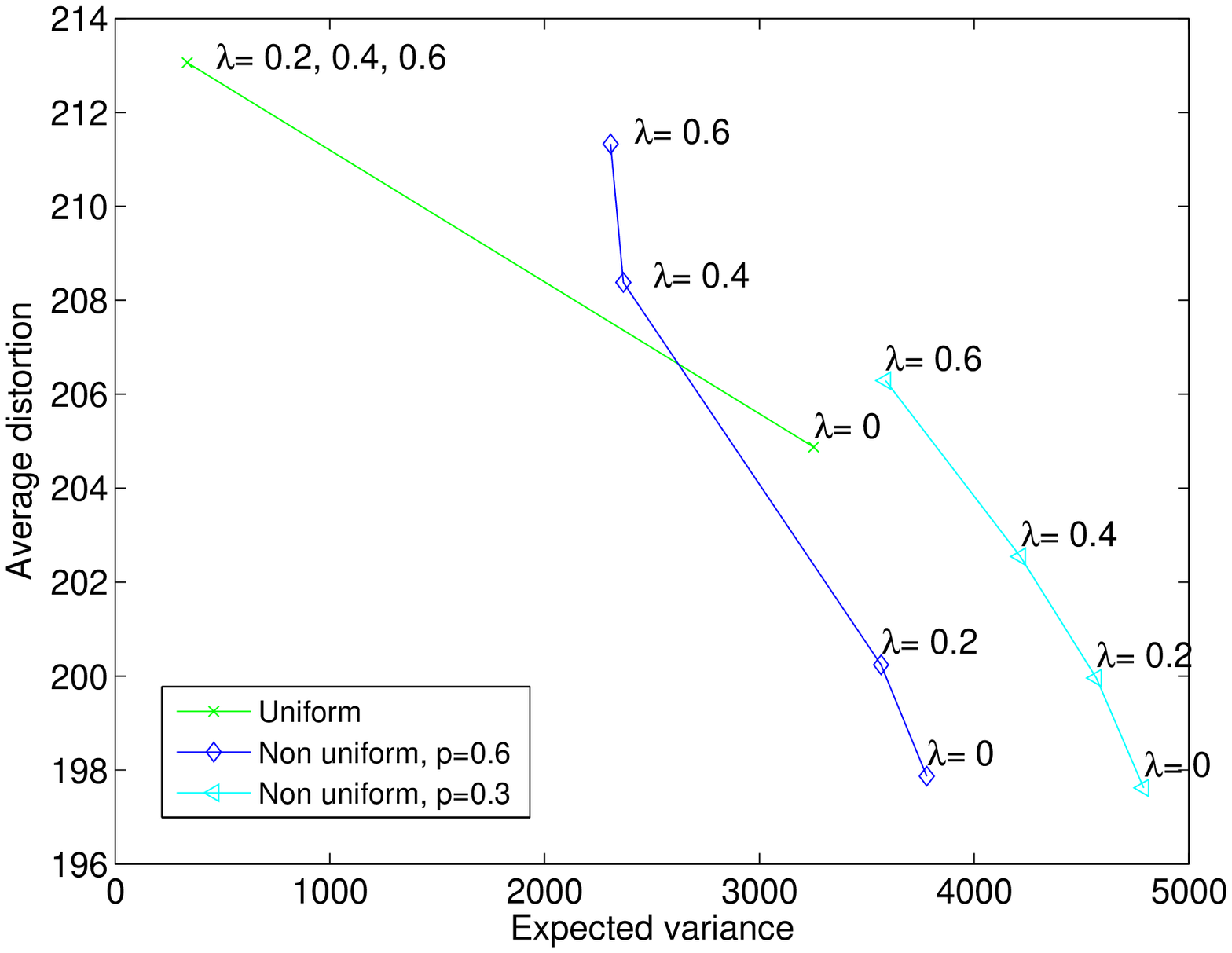} }  \hfill   
\subfigure[Directional navigation paths]{
\includegraphics[width=0.46\linewidth,  draft=false]{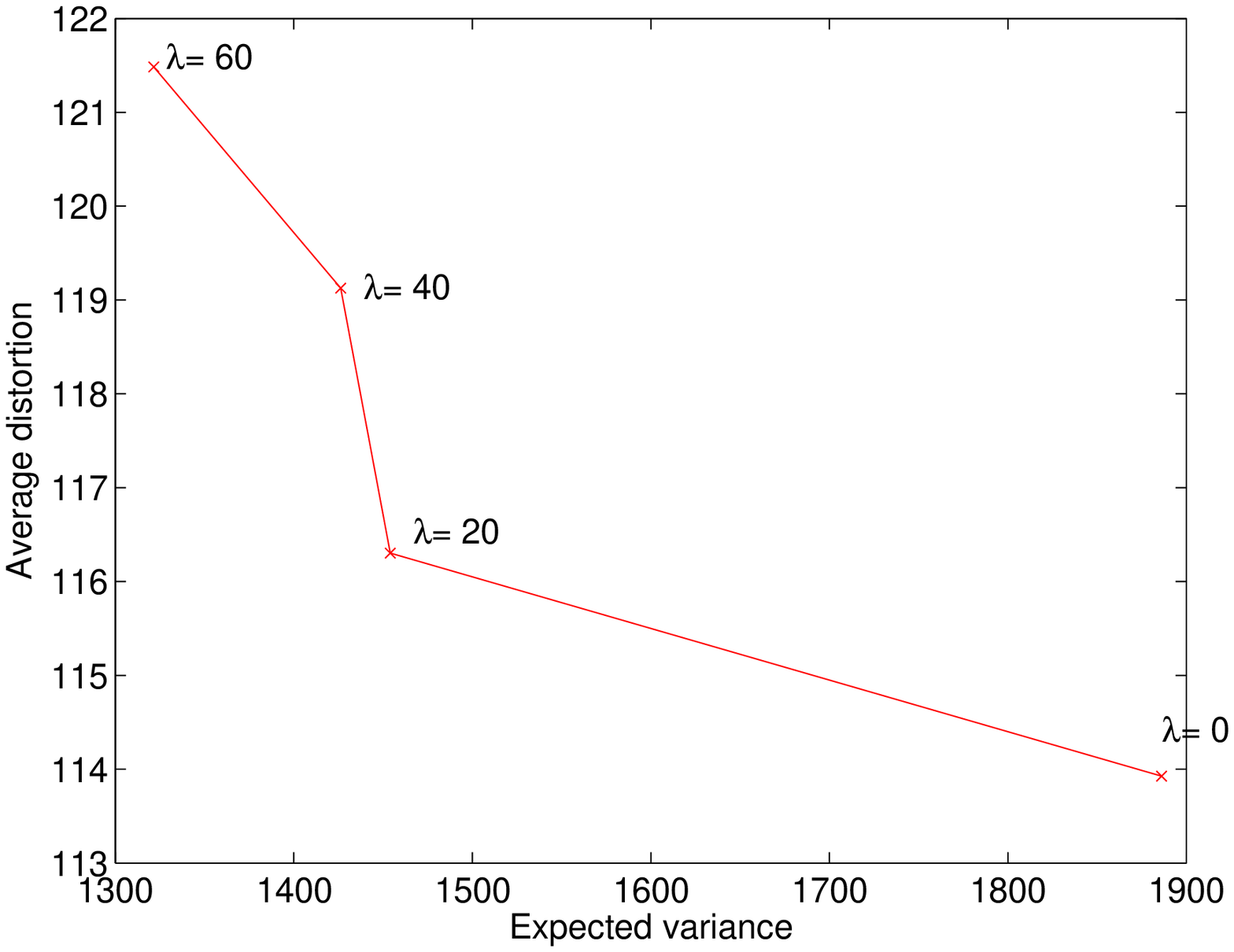} }
   \caption{Average distortion vs expected variance for the Synthetic sequence ($T_A=4$, $T_D=1$,   $\rho_S=2$,   $\rho_T=1$, static channel and dynamic navigation). } \label{fig:Synth_UniformNavigation_Tc1_Ts4}
\end{figure}

To give more intuitions on the distortion-variance tradeoff in different challenging scenarios, we now show the behavior of different navigation paths. In particular, we consider a \emph{uniform navigation}, where each user have the same probability of displaying the current view, or switching to the left or  right view. In this case the   camera popularity is $1/M$ for all views at each time instant.    We then consider a  \emph{non-uniform navigation}, where each user has a probability $p$ of displaying the current view and $(1-p)/2$ of switching to     left or   right view. Finally, we denote by  \emph{directional  navigation} the dynamic navigation considered before and shown in Fig. \ref{fig:camera_popularity}.   We have simulated  these different navigation paths and observed the performance  have been simulated and carried out results are provided in Fig. \ref{fig:Synth_UniformNavigation_Tc1_Ts4} for the Synthetic sequence, with $T_A=4$, $T_D=1$,   $\rho_S=2$, $\rho_T=1$, and a static channel ($C=2$). We can observe that the uniform navigation has a constant distortion-variance point for $\lambda>0$. Moreover, the directional navigation as well as the non-uniform  navigation with $p=0.6$ also has a limited reduction of the mean variance when $\lambda$ ranges from $0.4$ to $0.6$. This is also given by the fact that a more directional navigation path reduces the degree of freedom in the optimization, since some views are clearly dominant in the possible switching   from interactive users. A larger gain with increasing $\lambda$ is observed for the non-uniform navigation with $p=0.3$, where there is more randomness about users' interactivity.

   \begin{figure}
  \begin{center}
 \subfigure[Ballet sequence, $C=1.5$]{
 \includegraphics[width=0.45\linewidth,  draft=false]{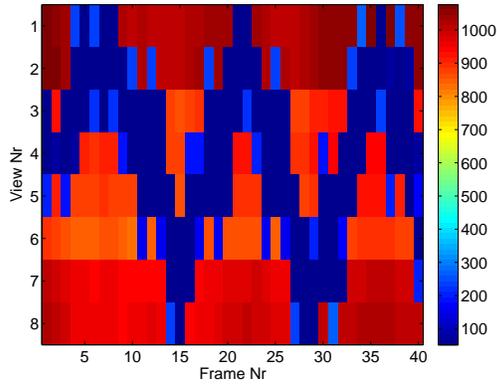} }  \hfill   
 \subfigure[Synthetic sequence, $C=3$]{
\includegraphics[width=0.45\linewidth,  draft=false]{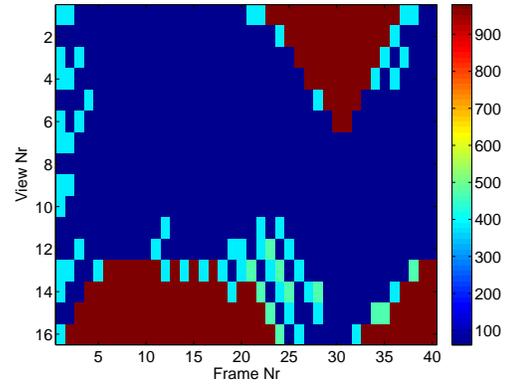}  } 
  \caption{Distortion experienced per  image, for each view and each time instant   ($T_A=4$, $T_D=1$,   $\rho_S=2$,  $\rho_T=1$, $\lambda=0.6$, static channel and dynamic directional navigation).  } \label{fig:ViewDistortion_2D}
  \end{center}
  \end{figure}   
  
\begin{figure}
\subfigure[Model-based PSNR]{
\includegraphics[width=0.46\linewidth,  draft=false]{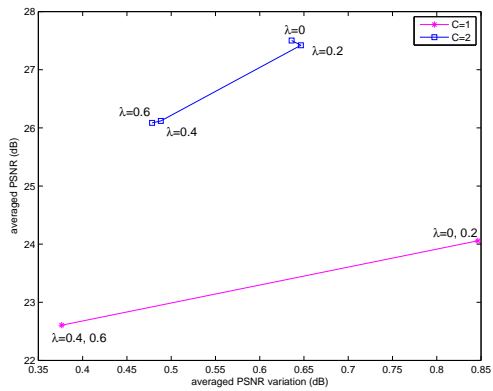} }  \hfill   
\subfigure[Experimental PSNR]{
\includegraphics[width=0.46\linewidth,  draft=false]{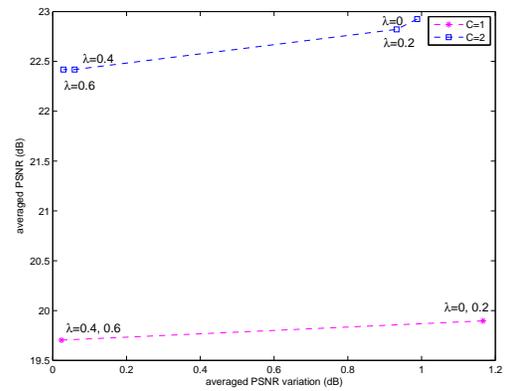} }   \caption{Average quality vs expected quality variance for the Ballet sequence ($T_A=1$, $T_D=3$,   $\rho_S=4$,  $\rho_T=1$, static channel and static navigation path). } \label{fig:Ballet_rhoS4_rhoT1_C1-25}
\end{figure}

 Finally, in Fig. \ref{fig:ViewDistortion_2D}, we provide the distortion experienced at each image in views and time, to show how this distortion changes depending on the possible navigation paths.  Results are provided for both the Synthetic and Ballet sequences, with $T_A=4$, $T_D=1$,   $\rho_S=2$, $\rho_T=1$, $\lambda=0.6$, a  directional navigation path, and a static channel with $C=1.5$ and $C=3$ for Ballet and Synthetic sequences, respectively. We can see that   the lowest distortion region follows the zig-zag behavior of the camera popularity (depicted in Fig.  \ref{fig:camera_popularity}), as a consequence of the optimization of the  popularity-weighted distortion in our scheduling algorithm.

To conclude, we validate our results by comparing our model-based results with experimental results. Note that in the above model-based results, we evaluate the average distortion (or the associated PSNR) from the model in Eq. \eqref{eq:RD_final_2}, while in the experimental results, the distortion is evaluated after actual reconstruction of the Ballet sequence from the received frames. 
In  Fig. \ref{fig:Ballet_rhoS4_rhoT1_C1-25},    the distortion as a function of the mean variance is provided for the Ballet sequence and two different bandwidths when   $T_A=1$, $T_D=3$,   $\rho_S=4$,  $\rho_T=0$ and both the channel and navigation paths are static.  For both model-based and experimental results,  the lower the channel bandwidth the lower the quality, as expected since less  views can be scheduled at each transmission opportunity for smaller channels. More interestingly,     by increasing $\lambda$ up to $0.6$ we can minimize the expected quality variance  at the price of a  reduced  average quality. However, while we experience a substantial reduction of the quality variance, the penalty in terms of average quality is most of the time marginal for both the model based and experimental results. Furthermore, we observe  that the qualitative behavior of the model-based results is similar to the experimental ones, validating the model considered in our paper.

Finally, we note   that the experienced PSNR in the experimental results ranges between $19.5$ dB and $23$ dB, which are very low PSNR values. This is mainly due to the fact that the system is highly constrained  with very low bandwidth and while some images are
received at very low quality in favor of some other more important scheduled frames, as shown in  
Table \ref{table:meanPSNR_vs_MLP_Cvarying} and Table \ref{table:meanPSNR_vs_MLP}. Table \ref{table:meanPSNR_vs_MLP_Cvarying} compares the average PSNR to the PSNR experienced over the most likely path (MLP) for the Ballet sequence in the scenarios of $\rho_s=2, \rho_t=0$,   and  dynamic navigation path (directional navigation). Different  channel bandwidth  values are considered  in the case of static channel.  For all  values of bandwidth $C$, the MLP PSNR is always higher than the average one;  we  also see that, by relaxing the constraints imposed in the optimization (i.e., increasing the bandwidth), the quality increases. Finally, although fixing the optimization parameter  $\lambda=0.6$ reduces the mean PSNR with respect to  $\lambda=0$, the quality over the MLP is not necessarily penalized. This is a consequence of the fact that large  $\lambda$ values imposed in the optimization   leads to a scheduling strategy that reduces the oscillations and if possible maintain a constant (and high) quality value over the MLP. Similar conclusions can be carried out from Table \ref{table:meanPSNR_vs_MLP}, where different navigation paths have been considered.

\begin{table*}[t]
\begin{center}
\caption{Mean PSNR vs most likely path (MLP) PSNR for  Ballet sequence in the settings of $\rho_s=2, \rho_t=0$, static channel, and  dynamic navigation path (directional navigation). Experimental results.} \label{table:meanPSNR_vs_MLP_Cvarying}
\begin{tabular}{c|c|c||c|c||c|c|}  
\cline{2-7}
\multirow{2}{*}{}
&   \multicolumn{2}{c||}{$C=2$}  &   \multicolumn{2}{c||}{$C=2.5$}  &   \multicolumn{2}{c|}{$C=3$}  \\  \cline{2-7}
 & Mean PSNR& MLP  PSNR & Mean  PSNR & MLP  PSNR& Mean  PSNR& MLP PSNR \\
 \hline 
 \multicolumn{1}{|c|}{$\lambda=0$} & $25.9$ &  $29.3$ & $26.5 $ & $30.3$ & $ 29$ & $31$ 
 \\ \hline
 \multicolumn{1}{|c|}{$\lambda=0.6$} & $25.7$ & $29.2$ &  $26.3$ &  $30.9$ & $28.9$ & $31.4$
   \\ \hline   
\end{tabular}
\end{center}
\end{table*}

%
%

\begin{table*}[t]
\begin{center}
\caption{Mean PSNR vs MLP PSNR for  Ballet sequence in the settings of $\rho_s=4, \rho_t=1$, static channel ($C=2$), and  dynamic navigation path. Experimental results.}\label{table:meanPSNR_vs_MLP}
\begin{tabular}{c|c|c||c|c|}  
\cline{2-5}
\multirow{2}{*}{}
 &        \multicolumn{2}{c||}{Non uniform Nav. ($p=0.6$)}  &   \multicolumn{2}{c|}{Uniform Nav. }  \\  \cline{2-5}
 & Mean PSNR& MLP  PSNR & Mean  PSNR & MLP  PSNR \\
 \hline 
 \multicolumn{1}{|c|}{$\lambda=0$} &  {$25.4$} & {$26.4$}	& {25.4} &  $26.4$ \\ \hline
 \multicolumn{1}{|c|}{$\lambda=0.6$} &  {$25.7$} & {$27.8$} & \centering{$25.9$} &  $27.3$    \\ \hline   
\end{tabular}
\end{center}
\end{table*}


\section{Conclusions}
\label{sec:conclusion}
We have investigated coding and scheduling strategies of redundant correlated sources in a multicamera system. 
 In particular, we have proposed  a novel rate-distortion model able to take into account the correlation level among cameras for different coding structures. Based on this rate-distortion function, we have proposed a dynamic packet scheduling algorithm, which opportunistically optimizes the transmission policy based on the channel capacity and source correlation. The best scheduling policy minimizes the popularity-weighted distortion 
while also reducing the distortion variations along most likely navigation paths experienced by  potential interactive users.   Because of the reward and coding dependency that subsists among frames, conventional solving methods cannot be adopted in our work. We have then proposed a novel trellis-based solving method that is able to decouple dependent and independent DUs in the trellis construction. This allows to reduce the computational complexity while preserving the optimality of the scheduling policy. Simulation results have demonstrated the gain of the proposed method compared to classical resource allocation techniques. This gain is due to the ability of the proposed algorithm to dynamically adapt the transmission strategy (and the coding structure accordingly)  to both the level of correlation experienced by each camera and the interactivity level experienced by potential users.  We have also shown that the proposed scheduling optimization is able to reduce the variations over the navigation path when the objective function is appropriately  designed.

\bibliographystyle{IEEEtran}
\bibliography{DSC}
%
%
%
%
%
%
%
%

\end{document}